\documentclass[12pt]{iopart}

\usepackage{color}
\usepackage{graphicx}
\usepackage{iopams}
\expandafter\let\csname equation*\endcsname\relax
\expandafter\let\csname endequation*\endcsname\relax
\usepackage{amsmath}
\usepackage{amssymb}
\usepackage[utf8]{inputenc}
\usepackage{url}
\usepackage{textcomp}

\begin{document}

\newcommand{\Comment}[1]{\textcolor{blue}{#1}}
\newcommand{\CommentImp}[1]{\textcolor{red}{#1}}
\newcommand{\ToDo}[1]{\textcolor{green}{(\textbf{ToDo:} #1)}}

\title[Quantum Walks with Dynamical Control]{Quantum Walks with Dynamical Control: Graph Engineering, Initial State Preparation and State Transfer}

\author{Thomas~Nitsche$^{1,*}$, Fabian~Elster$^{1}$,  Jaroslav~Novotn\'y$^{2}$, Aur\'el~G\'abris$^{2,3}$, Igor~Jex$^{2}$, Sonja~Barkhofen$^{1}$, Christine Silberhorn$^{1}$}
\address{$^1$ Applied Physics, University of Paderborn, Warburger Str. 100, 33098 Paderborn, Germany}
\address{$^2$Department of Physics, Faculty of Nuclear Sciences and Physical Engineering, Czech Technical University in Prague, B{\v r}ehov\'a 7, 11519 Prague, Czech Republic}
\address{$^3$Department of Theoretical Physics, University of Szeged, Tisza Lajos k\"or\'ut 84, H-6720 Szeged, Hungary}
\ead{*: tnitsche@mail.upb.de} 

\date{\today}

\begin{abstract}
Quantum walks are a well-established model for the study of coherent transport phenomena and provide a universal platform in quantum information theory.
Dynamically influencing the walker's evolution gives a high degree of flexibility for studying various applications.
Here, we present time-multiplexed finite quantum walks of variable size, the preparation of non-localized input states and their dynamical evolution. 
As a further application, we implement a state transfer scheme for an arbitrary input state to two different output modes.
The presented experiments rely on the full dynamical control of a time-multiplexed quantum walk, which includes adjustable coin operation as well as the possibility to flexibly configure the underlying graph structures. 
\end{abstract}

\maketitle

\section{Introduction}


The well-established concept of quantum walks \cite{aharonov_quantum_1993, kempe_quantum_2003} lays the foundation for the study of a rich variety of phenomena such as transport dynamics \cite{plenio_dephasing-assisted_2008, mohseni_environment-assisted_2008, ahlbrecht_molecular_2012, shikano_discrete-time_2014}, topological phases \cite{kitagawa_exploring_2010, kitagawa_topological_2012, asboth_symmetries_2012} or quantum computation \cite{lovett_universal_2010, childs_universal_2009, childs_universal_2013}. 
Current experimental implementations include nuclear magnetic resonance \cite{du_experimental_2003, ryan_experimental_2005}, trapped ions \cite{schmitz_quantum_2009,zahringer_realization_2010}, atoms \cite{karski_quantum_2009, genske_electric_2013}, photonic systems \cite{bouwmeester_optical_1999, do_experimental_2005, broome_discrete_2010, regensburger_photon_2011, cardano_quantum_2015, xue_observation_2014} and waveguides \cite{perets_realization_2008, bromberg_quantum_2009, peruzzo_quantum_2010, owens_two-photon_2011, sansoni_two-particle_2012, di_giuseppe_einstein-podolsky-rosen_2013, crespi_anderson_2013, meinecke_coherent_2013, poulios_quantum_2014}.


Time-multiplexed quantum walks have been introduced in 2010 with the first demonstration of a photonic quantum walk on a line \cite{schreiber_photons_2010}, which was the first experimental implementation of a coined discrete quantum walk.
Its advantage of requiring little resources while at the same time offering a high homogeneity and long-time stability against decoherence due to interferometric stability established a new experimental platform for studying complex coherence phenomena.
By introducing a fast switching electro-optic modulator the static coin operation was extended to a dynamical coin in \cite{schreiber_decoherence_2011}, which enabled the investigation of random perturbations with dynamically changing coins. This experiment succeeded in demonstrating the quantum to classical crossover by destroying the interferences, and showed Anderson localisation in a quantum walk system.

In the next step, the concept of time-multiplexed quantum walks was extended to the second dimension with the first experiment of a quantum walk on a 2-dimensional lattice \cite{schreiber_2d_2012}. 
Since this situation is mathematically equivalent to two walkers in one dimension, this system can be used as a highly controllable simulator for various 2-particle interactions in one dimension.
Still, all of these implementations have been realized on a static underlying graph and the principle of time-multiplexing appeared to be incompatible with reconfigurations  of the lattice structure.
In order to overcome this restriction, a sophisticated double-step scheme was invented, which allowed the demonstration of a quantum walk on a dynamical percolation graph \cite{elster_quantum_2015}.
Such a graph is defined as having its edges probabilistically added or removed in time to imitate porous or fluctuating media.
Using a quantum walk on a dynamical percolation graph, subtle decoherence effects in presence of randomly distributed, fluctuating gap defects were studied and clear signatures of non-Markovian behaviour were observed.

In this article, we present three new applications to further emphasize the capabilities and the flexibility of our current time-multiplexed quantum walk setup:
First, we show the walker's time evolution on finite graphs with variable sizes.
Second, we exploit the dynamical coin to prepare for the first time in-situ non-localised initial states with controlled polarization and well-defined phase.
By comparing their evolution, their strong dependency on the input polarization is revealed.
And third, we demonstrate a state transfer scheme as suggested in \cite{zhan_perfect_2014} and route an arbitrary and unknown initial state to two different detector positions achieving high fidelities between input and output state up to 99.4\%.
We find the predicted periodicity in the time dynamics of the walker localizing at the original position \cite{zhan_perfect_2014}.

The article is structured as follows:
It starts with a conceptual introduction and a review of time-multiplexed quantum walks with dynamical coin control in section \ref{sec:concepts}, including the description of time-multiplexing and the fundamentals on dynamical coin control. 
Then we present the experimental implementation of the quantum walk with a dynamically controlled coin in section \ref{sec:exp_imp}. 
Here, we also provide details on experimental challenges and difficulties inherent in the nature of time-multiplexed systems and give insights into the realised solutions.
These concern the concrete timings, how to handle the different power levels and how to deal with the ubiquitous losses. 
Section \ref{sec:results} is dedicated to the presentation of the experimental results of the three applications of the dynamical coin: the implementation of a finite walk, in-situ preparation of non-localised initial states and the demonstration of state transfer schemes. 
Finally, we conclude in section \ref{sec:conclusion}.

\section{Concepts}
\label{sec:concepts}
\subsection{Fundamentals} 
\label{subsec:fundamentals}
The concept of quantum walks merges the classical model of the random walk with the quantum mechanical description of a system via its wave function. 
From the classical random walk originates the idea that the walker's path is determined by a coin toss. 
While the walker takes one of the possible paths in a classical random walk, in a quantum walk the amplitude of the wave function is split up into components taking the different paths. 
This results in the superposition of components originating from different paths and consequently interference effects are observed. 
\\
The state ${| \Psi \rangle}$ of the quantum walker is described by a vector in a tensor product space $\mathcal{H}_x\otimes\mathcal{H}_c$ of a position Hilbert space $\mathcal H_x$ and a coin Hilbert space $\mathcal H_c$.
Let $| x \rangle $ and $| c \rangle $ denote the basis vectors of the position and the coin space then we can write the full wave function at time $t$ as:

\begin{equation}
| \Psi (t) \rangle =  \sum_x \sum_c a_{x,c}(t)| x \rangle \otimes | c\rangle ~.
\label{eq:Psi}
\end{equation}
with the time-dependent amplitudes $a_{x,c}(t) \in \mathbb C$.

%

Assuming a time-independent effective Hamiltonian, the evolution of the walker's wave function with the number $n$ of discrete time steps is given by the following expression:

\begin{eqnarray}
 |\Psi(t_n) \rangle = e^{- i n \cdot \hat{H}} |\Psi(t_0) \rangle ~.
\end{eqnarray}

In the discrete time quantum walk (DTQW), the evolution unitary $\hat{U}$ for one discrete step consists of a coin operation $\hat{C}$, the analog of the classical coin toss, acting only on the internal state and a subsequent step operation $\hat{S}$ modifying the external state:

\begin{equation}
|\Psi(t_1) \rangle = \hat{U} |\Psi(t_0) \rangle = \hat{S} \hat{C} ~|\Psi(t_0) \rangle ~.
\label{eq:SC}
\end{equation}
For our realisation of a quantum walk on a 1D grid, the polarisation, expressed in the basis vectors $| H \rangle = (1, 0)^{\mathrm{T}}$ and $| V \rangle = (0, 1)^{\mathrm{T}}$, represents the internal state, which then determines the operation $\hat{S}$ on the positions $x\in \mathbb Z$:

\begin{eqnarray}
\hat{S} = \sum_{x \in \mathbb Z} \left( | x + 1, H \rangle \langle x, H |  + | x - 1, V \rangle \langle x, V |\right) ~.
\label{eq:Step}
\end{eqnarray}

The part of the walker's wave packet in horizontal polarisation $| H \rangle$ undergoes an increase of position by one, while the part in vertical polarisation $ | V \rangle$ has the position decreased by one. 
In general the step operator $\hat{S}$ carries the information about the possible paths for the walker, i.e. the pattern of connections (edges) between possible positions (here: a 1-d line) given by the underlying graph.
More complex graph configurations, e.g. a line with missing links or boundaries, thus require a modified step operator compared to expression (\ref{eq:Step}).

\subsection{Photonic Quantum Walks with Time-Multiplexing} 
\label{subsec:multiplex}
\begin{figure}[t]
	\centering 
		\includegraphics[width=\columnwidth]{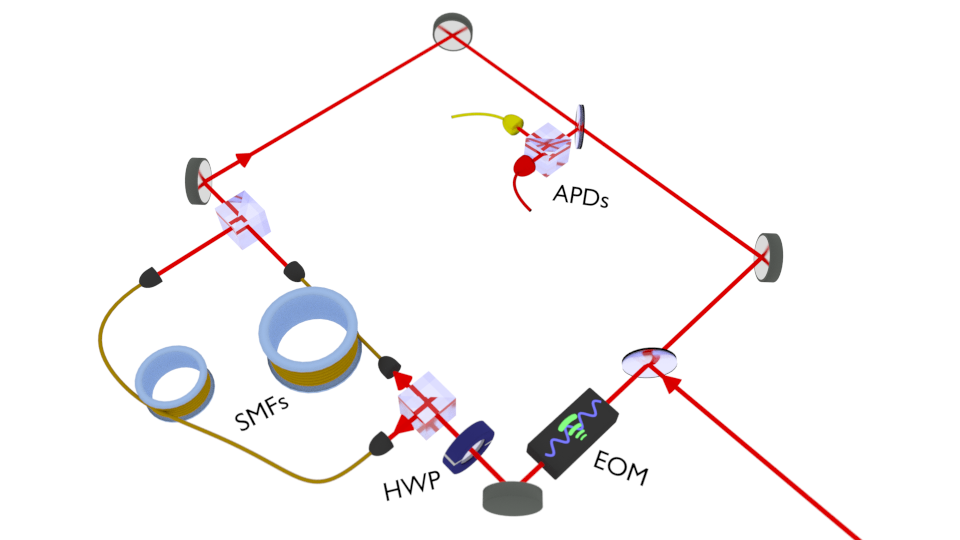}
	\caption{Schematic of our implementation of a time-multiplexed quantum walk with an electro-optic modulator (EOM) and a half-wave plate (HWP) (here exemplarily) carrying out the coin operation, two single mode fibres (SMFs) introducing the time delay and the avalanche photo-diodes (APDs) used for the read-out of the walker's state}
	\label{fig:Setup}
\end{figure}

We realise the DTQW with a single photonic walker and use its polarisation as the coin state.
Implementing the quantum walk with the position as the external degree of freedom, however, limits the scalability of the setup (for the example of a Galton board-like beam splitter cascade the number of components increases quadratically with the number of steps).
Additionally, decoherence effects will occur due to experimental inaccuracies of the growing number of components. 
Thus, we take a different approach and apply a loop geometry \cite{schreiber_photons_2010}: 
The concept of time-multiplexing is based on encoding the external state of the quantum walker into the time domain  (see Figure \ref{fig:Setup}). 
Initialising a photonic quantum walk, a coherent laser pulse is sent into the setup with a given input polarisation. 
First, the pulse passes standard wave plates or a fast switching electro-optic modulator (EOM), which realise the desired coin operation (see sec. \ref{subsec:graph_implement} for details).
Then, a polarisation dependent splitting of the pulse is carried out by a polarising beam splitter (PBS).
By routing the pulses afterwards through single-mode fibres (SMF) of different lengths we introduce a well-defined time delay between them.
This splitting operation in combination with the time delay and the subsequent merging of the paths at the second PBS constitutes one shift of the time-multiplexed quantum walk according to eq. \ref{eq:Step}. 
We interpret the part arriving earlier as having undergone a reduction of the position by one and the component arriving later as having been subjected to an increase of the position by one.
Hence, every spatial position is uniquely represented by its arrival time.

\begin{figure}[t]
	\centering 
		\includegraphics[width=\columnwidth]{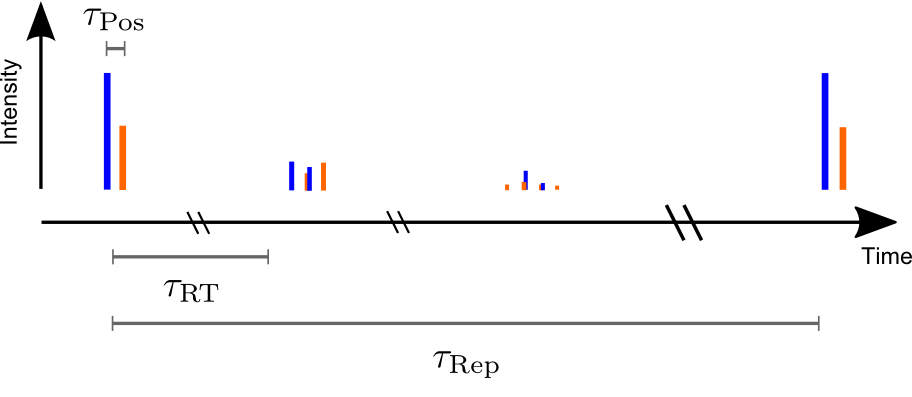}
	\caption{Schematic illustration of the 3 relevant timings (not to scale): the position spacing $\tau_\mathrm{Pos}$, the roundtrip spacing $\tau_\mathrm{RT}$ and the time $\tau_\mathrm{Rep}$ between two consecutive experiments defining the repetition rate; counts for vertically polarized light are represented by blue bars and counts for horizontally polarized light by orange bars (schematically)}
	\label{fig:Timings_Sketch}
\end{figure}

As we want to achieve the $n$-times application of coin and the step operation, the output is again fed into the setup, resulting in a loop-like architecture of the experimental implementation.
In every step a small amount of intensity is coupled out and sent to the detection unit consisting of another PBS and two avalanche photodiodes (APDs) allowing for polarisation resolved measurements of the walker's time evolution.
\\
Figure \ref{fig:Timings_Sketch} illustrates the 3 relevant timings in the time-multiplexed quantum walk: The splitting of the two polarisation components introduces a delay of $\tau_\mathrm{Pos}$ between them given by the length difference between the two SMFs.
The roundtrip time $\tau_\mathrm{RT}$ is defined by the length of the shorter fibre (plus the free-space parts) and denotes the step separation.
Two subsequent experiments are delayed by $\tau_\mathrm{Rep}$ which must be set long enough to observe the desired number of steps without overlapping.\\
The proper tailoring of these timings is essential for the time-multiplexing scheme and we dedicate sec. \ref{subsec:Timings} to its experimental details.

\subsection{Dynamical coin and dynamical graph} 
\label{subsec:graph_implement}
\begin{figure}[t]
	\centering 
		\includegraphics[width=10cm]{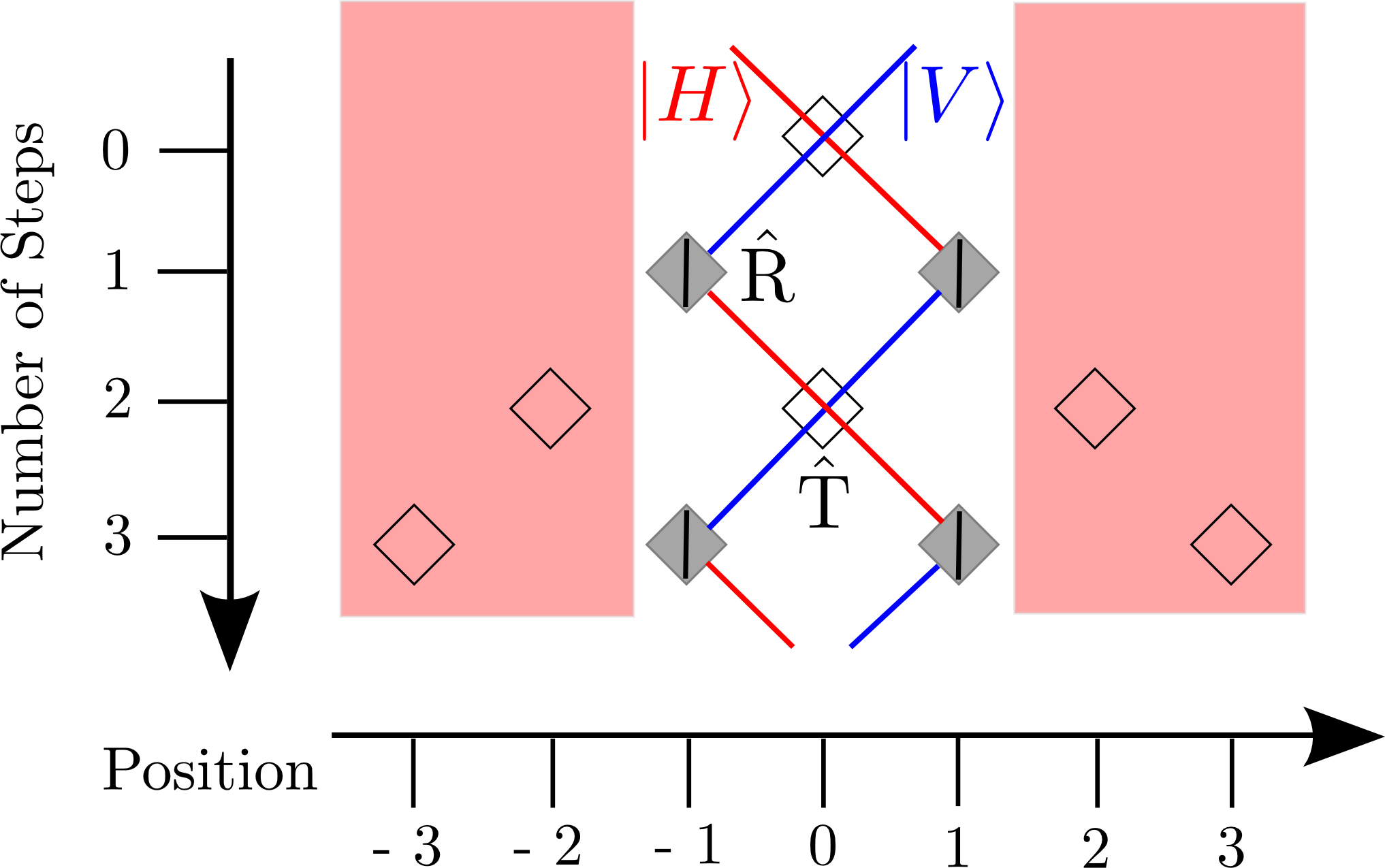}
	\caption{Illustration of how the spread of the walker's external state is limited via polarisation switching; a reflection $\hat{R}$ is represented by a grey box and switches the horizontal (red lines) into vertical (blue lines) polarization and vice versa. The transmission operation $\hat{T}$ is represented by a white box leaving the polarizations unchanged. Note that in the setup the spatial position is in the experiment mapped into the time domain.}
	\label{fig:Kaskade_Timings}
\end{figure}

As described above a quantum walk includes a coin operation acting on the internal degree of freedom, in our case the polarisation.
In the experiment the polarisation can be statically rotated in a straight-forward way by applying retardation plates. 
For example a typical coin, implemented by a half-wave plate (HWP) at an angle of $22.5^\circ$, is the Hadamard coin, which transforms an input comprising one polarization into a fair superposition of both polarizations. 
Another balanced coin can be implemented by a quarter-wave plate (QWP) at an angle of $45^\circ$ (see subsection \ref{subsec:dyna_coin}).
Quantum walks having the spatial position as the external degree of freedom can rely on static phase-shifters in order to implement a position-dependent coin operation \cite{xue_trapping_2014}. In a time-multiplexed setup in a loop architecture, however, such devices can only realise operations which are the same for all steps and positions.
\\
Introducing a dynamical coin dramatically increases the versatility of a quantum walk as a model system: It offers new possibilities, e.g. in altering the structure of the underlying graph \cite{elster_quantum_2015} (see also subsection \ref{subsec:finite}) as well as in ``in-situ'' preparation of the initial state of the quantum walk (see subsection \ref{subsec:state_prep}). 
\\
A direct manipulation of the graph structure would mean to modify the step operator which carries the information on the walker's possible paths.
Since in our setup the step operation is realised by a fixed combination of PBSs and SMFs this can obviously \emph{not} be dynamically changed.
However, the displacement direction in the step operation depends on the polarisation (see equation \ref{eq:Step}) and the polarization can be addressed dynamically via fast switching EOMs.
Thus a clever manipulation of the polarization allows to prevent that intensity from a certain position is shifted in a certain direction (e.g. to the right), which is the same effect as breaking the respective link (in this case the right one) would have.
\\
We will present here in more detail how the dynamical coin is harnessed for the implementation of a quantum walk on a finite graph:
Figure \ref{fig:Kaskade_Timings} shows in a Galton board-like scheme a possible evolution of the walker using two specific coin operations, the reflection and the transmission. 
A reflection operation $\hat{R}$ (grey box) denotes a polarization switching from $| H \rangle$ to $| V \rangle$ and vice versa.
Leaving the polarisations unchanged corresponds to the transmission operation $\hat{T}$ (white box). 
In the given example, we intend to limit the spread of the walker's external state to the positions $-1,0,1$, thus it must not evolve into the red areas. 
This corresponds to a reflection operator at $x= -1$ for the left boundary and at $x = 1$ for the right boundary.
\\
The example illustrates on the one hand that time-multiplexing techniques in combination with dynamical coin control provide a powerful tool for implementing not only dynamical coin operations but also non-trivial graph structures. 
On the other hand, it also shows the significance of properly timed polarisation switchings, which require accurate engineering of the hardware and the software of the setup. 
The presentation of its details will be given in the following sections.

\subsection{Implementation of dynamical coin operations via an EOM}	 
\label{subsec:dyna_coin}
For the implementation of non-trivial graphs a device is needed that switches between reflection and transmission operations $\hat{R}$, $\hat{T}$ and, possibly, a third mixing coin $\hat{C}$.
In order to address single positions the device must exhibit a switching speed that is comparable to the position separation $\tau_\mathrm{Pos}$ (here: 46.5\,ns).

\begin{figure}[t]
	\centering 
		\includegraphics[width=10cm]{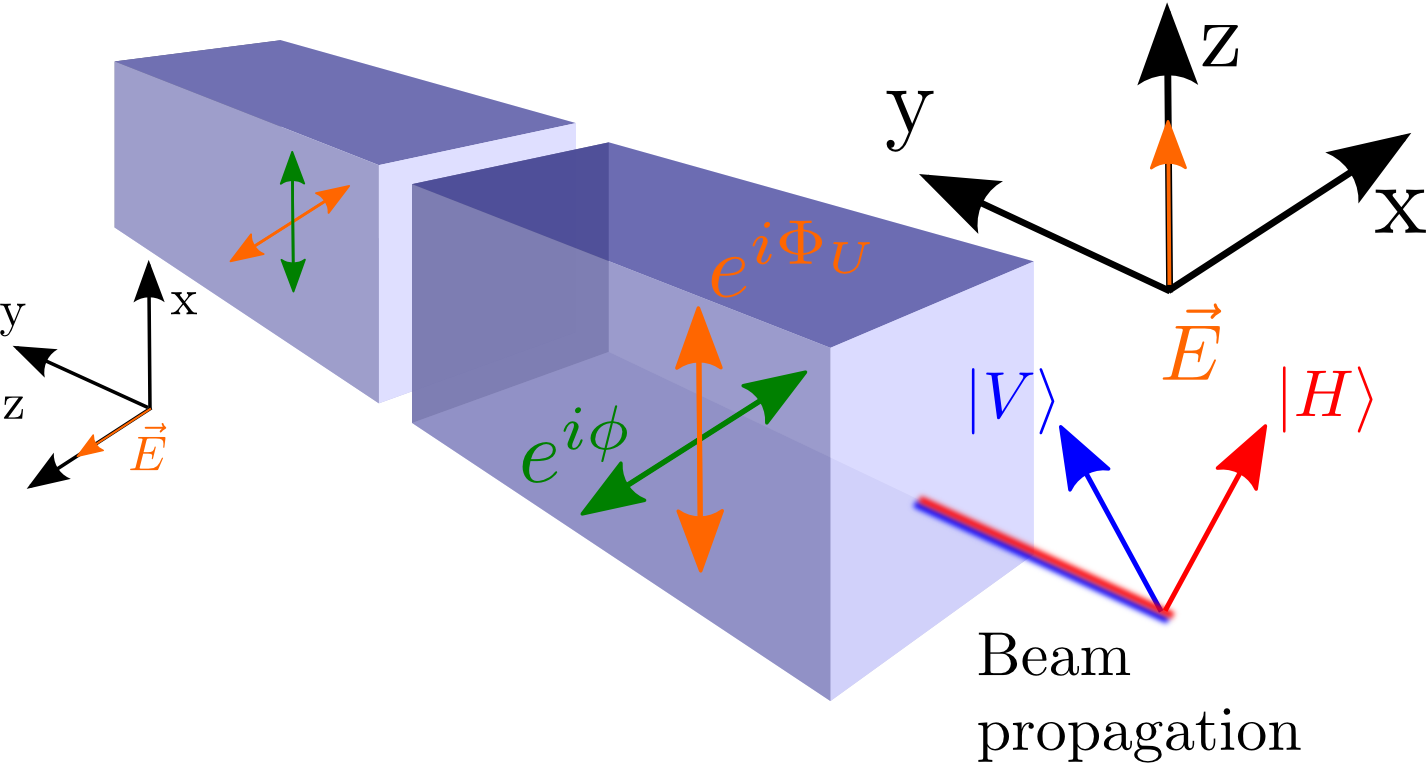}
	\caption{Schematic of the two EOM crystals; the reference frame is given by the orientation of the crystal axes and is rotated by 45$^\circ$ with respect to $| H \rangle$ and $| V \rangle$: the phase $\phi$ originating from the natural birefringence is accumulated along the $x$-axes, while the phase $\Phi_\mathrm{U}$ originating from the application of $\vec{E}$ is accumulated along the $z$-axes of the crystals.}
\label{fig:EOM}
\end{figure}

In order to meet the challenging requirements on homogeneity, accuracy of rotation angle and switching speed, we thus apply an electro-optic modulator (EOM, see Figure \ref{fig:EOM}).
This device operates based on the Pockels effect to manipulate the polarization of light passing through it: The refractive index of a medium is changed along the direction of an applied voltage, thus introducing or altering birefringence, which can be done at high speeds.
The Pockels cell used in the setup consists of a rubidium titanyl phosphate (RTP) crystal that exhibits natural birefringence.
Since we need to set the actually realised operation by switching on or off the external voltage, the natural birefringence has to be compensated for: This was achieved by cutting the crystal in two halves and rotate them against each other by 90$^\circ$.

In order to obtain the desired reflection and transmission matrices, aligning the crystal to an rotation angle of 45$^\circ$ with respect to the H- and V-axes is crucial.
Still it is not trivial which concrete operators can be realised with this device, since even phases are significant because the interference patterns are strongly phase-sensitive.
As the external voltage is applied along the $z$-axes of the crystals we can derive the expression for the rotation operation of the EOM in the H-V-basis:
\begin{center}
\begin{equation}
\begin{split}
\hat{C}_\mathrm{EOM}(U) &= 
R(45 ^\circ) \cdot \hat{C}_\mathrm{crystal~1} \cdot \hat{C}_\mathrm{crystal~2} \cdot R(-45 ^\circ)
\\
\\
&= \frac{1}{\sqrt{2}}
\begin{pmatrix}
1 & -1 \\
1 & 1 \\
\end{pmatrix} 
\begin{pmatrix}
e^{i \Phi_U} & 0 \\
0 & e^{i \varphi} \\
\end{pmatrix} 
\begin{pmatrix}
e^{i \varphi} & 0 \\
0 & e^{-i \Phi_U} \\
\end{pmatrix} 
\frac{1}{\sqrt{2}}
\begin{pmatrix}
1 & 1 \\
-1 & 1 \\
\end{pmatrix} 
\\
&= e^{i \varphi} \cdot
\begin{pmatrix}
\cos(\Phi_U) & i \sin(\Phi_U) \\
i \sin(\Phi_U) & \cos(\Phi_U) \\
\end{pmatrix} .
\end{split}
\label{eq:C_EOM_U}
\end{equation}
\end{center}
where $R( \pm 45 ^\circ)$ are the rotation matrices  for the basis transform, $\hat{C}_\mathrm{crystal~1}$ and $\hat{C}_\mathrm{crystal~2}$ represent the operation in the respective halves of the crystal and $\Phi_\mathrm{U}$ the additional phase induced by the Pockels effect.
In principle all operators of this shape can be realised, just by setting an appropriate voltage to attain the needed phase $\Phi_\mathrm{U}$.
However, the entries of the matrices either assume merely real or imaginary values, so that the possible operations do not exhaust the full space of unitary $2\times2$ matrices and we have to stay within this constraint regarding the phases when choosing coin operations. In the coherent evolution of the walker these phase terms are of fundamental significance.

For the modification of the graph structure we are interested in 3 special cases:
For $\Phi_U = 0$ the EOM realises the transmission operator $\hat{T} = \mathbf 1$ when setting the global phase $\varphi = 0$.
For an appropriate choice of $U$ yielding $\Phi_U = \frac{\pi}{2}$ we obtain the reflection operator $\hat{R}$ with $i$-phases on the off-diagonals. 
Setting $\Phi_U$ to $\frac{\pi}{4}$ realises a balanced coin $\hat{C}_\mathrm{QWP}$ that transforms e.g. $| H \rangle$ into an equal superposition of both polarisations and is equivalent to a quarter-wave plate (QWP):

\begin{equation}
\hat{T} = 
\left(
\begin{array}{cc}
1 & 0 \\
0 & 1\\
\end{array}
\right),~~~~ 
\hat{R} = 
\left(
\begin{array}{cc}
0 & i \\
i & 0 \\
\end{array}
\right),~~~~
\hat{C}_\mathrm{QWP} = \frac{1}{\sqrt{2}}
\left(
\begin{array}{cc}
1 & i \\
i & 1 \\
\end{array}
\right).
\label{eq:C_EOM_T}
\end{equation}

\begin{figure}[t]
	\centering 
		\includegraphics[width=10cm]{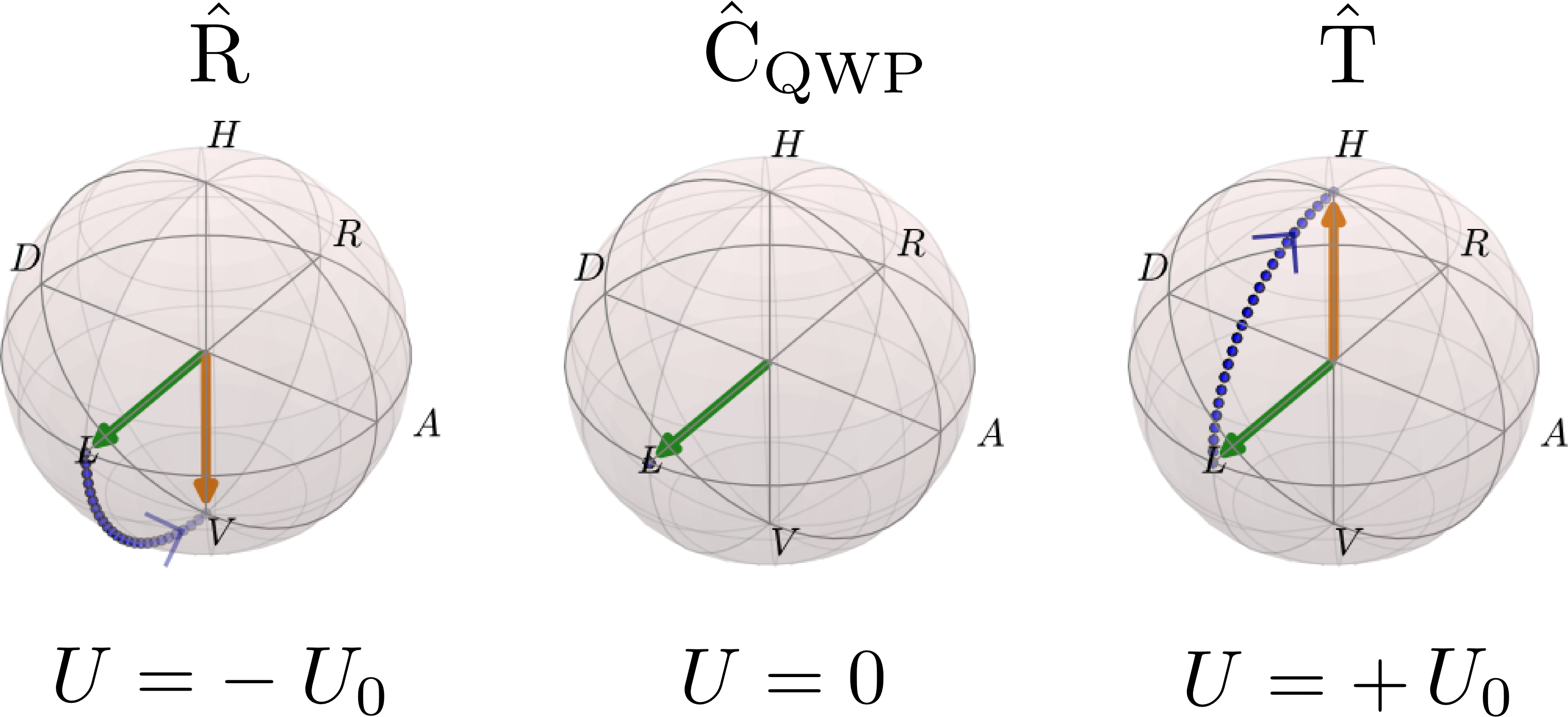}
	\caption{Bloch sphere representation of the EOM switching states corresponding to the three operations $\hat{R}$, $\hat{C}_\mathrm{QWP}$ and $\hat{T}$ exemplary for an horizontally polarised input state; a static QWP is placed in front of the EOM to realise $\hat{C}_\mathrm{QWP}$ in case no voltage is applied}
\label{fig:Bloch_too_fast}
\end{figure}

For our EOM in use there is another restriction: Within one experiment it is able to switch on or off a manually preset voltage with positive or negative sign, that means that we are limited to the three phases $\Phi_U = 0, \pm \Phi_{U_0}$ for one experiment.
In order to implement the three different operations anyhow, we place a static QWP with $\hat{C}_\mathrm{QWP}$ in front of it. 
Now we have a balanced coin when no voltage is applied in the EOM and can switch to reflection or transmission by applying a positive or negative voltage $U_0$, respectively (Fig. \ref{fig:Bloch_too_fast}).

\section{Experimental implementation} \label{sec:exp_imp}

\subsection{Timings} \label{subsec:Timings}
The time-multiplexing scheme is based on three different time scales which have to be designed appropriately in order to guarantee the unique mapping between arrival time and step and position assignment of a pulse.
In addition, technical restrictions must be taken into account, e.g the dead times of the detectors, the overall measurement time and where required the switching speed for the dynamical coin. 

The smallest time scale is the position separation $\tau_\mathrm{Pos}$ given by the runtime difference in the two fibres.
Since the walker's wave function is distributed over more and more positions with growing step number, the roundtrip time $\tau_\mathrm{RT}$ must be chosen long enough to prevent the overlapping of pulses belonging to different steps up to a maximal step $m$.
Such an unambiguous assignment of the positions requires that the latest position of the $n$-th step has no overlap with the earliest position of the $(n+1)$-th step (see Figure \ref{fig:Timings_Sketch}).
Since the number of possible positions of the $m$-th step equals $m + 1$, we find the following relation between the time scales:
\begin{equation}
(m+1) \cdot \tau_\mathrm{Pos} < \tau_\mathrm{RT}
\label{eq:Pos_space}
\end{equation}
In the presented setup we chose fibres of lengths 145\,m and 135\,m, respectively.
On the one hand, the resulting length difference of 10\,m leads to a position separation of $\tau_\mathrm{Pos} = 46.5$\,ns matching the deadtimes of the APDs and the switching speed of the EOM of around 50\,ns.
On the other hand, the spacing of the roundtrips $\tau_\mathrm{RT}$ is determined by the length of the shorter fibre (plus the free space parts), which corresponds to 685\,ns allowing for adding up 14 multiples of $\tau_\mathrm{Pos}$.
Thus, our setup allows for more than 13 steps only if explicit boundary conditions apply, restricting the walk to a finite number of positions.

The duration of one experimental run $\tau_\mathrm{Rep}$ is then given by the maximal step number times the roundtrip time.
In order to obtain reliable statistics for ensemble measurements, the experimental data is obtained via averaging over many runs of the experiment taking place with a repetition rate of $f_\mathrm{Rep} = \tau_\mathrm{Rep}^{-1}$.
Here, the entire duration of one experimental run of nearly 9\,$\mu$s allows for a repetition rate of $110$\,kHz.

For the finite walk (see sec. \ref{subsec:finite}) the limited number of populated positions prevents their overlapping, thus we can achieve step numbers up to 21 while running the experiment at 50\,kHz.
In both cases we get reliable statistics of the experimental data within a few minutes.

\subsection{Analysis of Photon Numbers for Reliable Detection} \label{subsec:numbers}

For several applications of quantum walks we aim for a reliable monitoring of the walker's full time evolution and not only for a final outcome distribution.
For achieving this goal we have to deal with several experimental challenges, e.g. a broad power range of pulses arriving at the detectors, reduction of multi-photon contributions at the APDs and accepting higher round trip losses when sending light to the detection unit in every step.
Hence a proper analysis of power levels and photon numbers is essential.

In order to track the walker's dynamics over time, a certain proportion of the light is coupled out in every step via probabilistic outcouplers and sent to the detetion unit. 
Similarly, the photons are initially fed into the setup via a probabilistic incoupler. 
The proper adjustment of the initial power level as well as the in- and outcoupling ratio, is critical for reliably measuring the walker's state for step numbers as high as possible: 
On the one hand, the intrinsic properties of ``click''-detectors demand for negligible multi-photon contributions in the steps to be analyzed.
On the other hand, we want to conduct measurements with an acceptable signal-to-noise ratio (SNR) for as high step numbers $n$ as possible.
However, due to the loop architecture the losses, including the outcoupled portions of the light, scale exponentially with $n$ limiting the maximal observable step number (for a detailed analysis of the occuring losses see next subsection). 
Consequently, a trade-off between a sufficient SNR for higher steps and a low enough initial energy to not damage the detectors and to reduce multi-photon contributions must be found.

\begin{figure}[t]
	\centering 
		\includegraphics[width=\columnwidth]{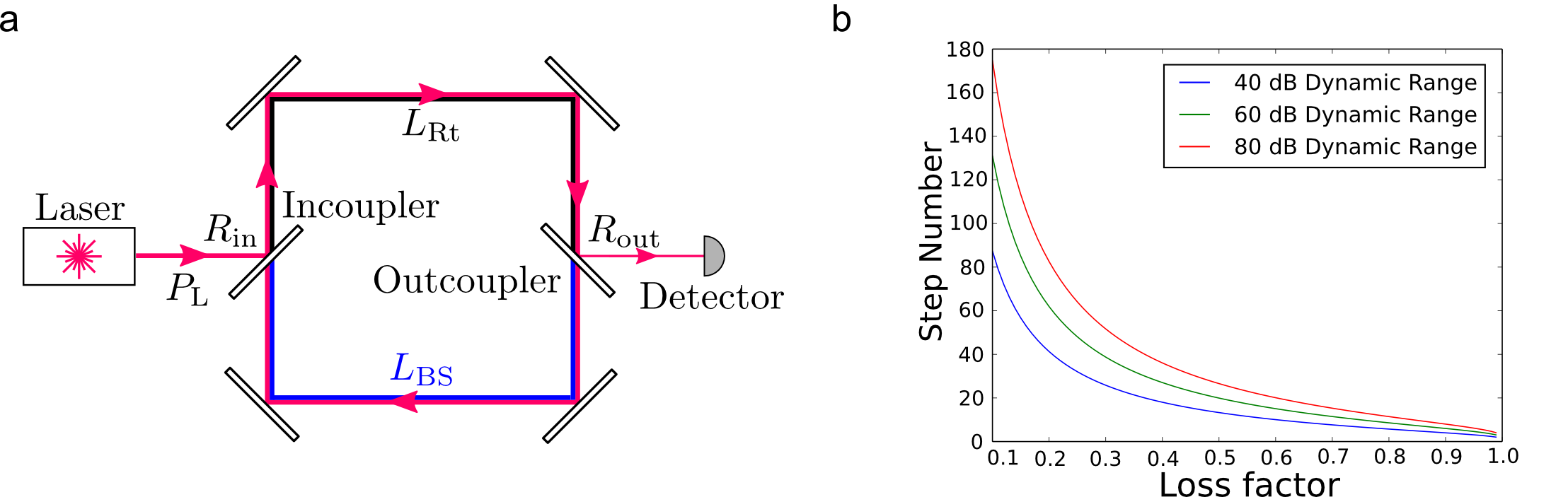}	
	\caption{Left figure: Illustration of the relevant quantities for determining $N_{\mathrm{phot}, n}$:  The roundtrip losses $L_{\mathrm{RT}}$ until the outcoupler are marked by the black line and the losses $L_{\mathrm{BS}}$ from the outcoupler to the beginning of a new roundtrip by the blue line, $R_\mathrm{in}$ and $R_\mathrm{out}$ mark the reflectivities of the in- respectively the outcoupler, $P_\mathrm{L}$ is the power of the incident laser pulses.
Right figure: The attainable step number depending on the loss factor for a roundtrip: the step numbers are given for a dynamic range of the detectors of 40 dB (blue curve), 60 dB (green curve) and 80 dB (red curve).}
\label{fig:roundtrip}
\end{figure}

In the following we perform an estimation for the photon number $N_{\mathrm{phot}, n}$ arriving at the detectors for a single coherent pulse depending on the roundtrip number $n$ (see Fig. \ref{fig:roundtrip} a) for an illustration of the considered quantities) in order to find the optimal power levels: 
The number of photons per incident laser pulse is given by the ratio of the initial pulse intensity before the incoupler $P_\mathrm{L}$ to the photon energy $E_\mathrm{phot}$ and the repetition rate $f_\mathrm{rep}$. 
The incoupler reflectivity $R_\mathrm{in}$ determines the proportion of these photons are coupled into the setup.
 Here, they are subject to $n$-times (number of roundtrips) the roundtrip loss factor $L_{\mathrm{RT}}$ until the outcoupler and $(n -1)$-times to the loss factor $L_{\mathrm{BS}}$ from the outcoupler to the beginning of a new roundtrip. 
Eventually, the incoupler reflectivity $R_\mathrm{in}$ gives the percentage of the remaining photons that are directed to the detectors. The factor $(n+1)^{-1}$ accounts for the number of positions the walker is split-up to in step $n$. 
When estimating an upper bound for the photon numbers, we assume that the power is concentrated at only one position and thus set this factor to 1:    
\begin{eqnarray}
N_{\mathrm{phot}, n}  = \underbrace{\frac{P_\mathrm{L}}{E_\mathrm{phot} \cdot f_\mathrm{rep}}}_{\scriptsize\begin{array}{l}\mathrm{photons \ per} \\ \mathrm{incident\ laserpulse}\end{array}} \cdot \underbrace{R_\mathrm{out} \cdot R_\mathrm{in}}_{\scriptsize\begin{array}{l}\mathrm{in- \ and} \\ \mathrm{outcoupling}\end{array}}  \cdot \underbrace{L_\mathrm{RT}^n \cdot L_\mathrm{BS}^{n -1}}_{\scriptsize\begin{array}{l}\mathrm{roundtrip} \\ \mathrm{losses}\end{array}} \cdot \underbrace{\frac{1}{(n + 1)}}_{\scriptsize\begin{array}{l}\mathrm{number \ of} \\ \mathrm{positions}\end{array}}
\label{eq:photon_number}
\end{eqnarray}
In a conservative estimation for the properties of our experiment, we consider a roundtrip loss factor $L_{\mathrm{RT}}$ until the outcoupler of 0.5, a loss factor $L_{\mathrm{BS}}$ from the outcoupler to the beginning of a new roundtrip of 0.97, in- and outcoupler reflectivities $R_\mathrm{in}$ and $R_\mathrm{out}$ of 2\textperthousand  ~respectively 7$\%$, a power level $P_\mathrm{L}$ before the incoupler of 1.67 $\cdot 10^{-9}$\,W, a photon energy $E_\mathrm{phot}$ of 2.46 $\cdot 10^{-19}$\,Ws for a wavelength of $805$\,nm and a repetition rate $f_\mathrm{rep}$ of $10^5$\,Hz. 

For single-photon APDs as used in our setup, a single photon generates a current peak (''click'') that is detected and counted as the arrival of at least one photon at the APD. Afterwards, APDs exhibit a dead-time of $\approx$ 50\,ns, which is slightly above the separation of the positions of 46.5\,ns. 
Any number of photons arriving within the dead-time-window is accounted for as just one ''click''. 
This may lead to errors in the measured occupation probabilities if there is a significant chance that several photons arrive at the same time.
An estimation according to eq. (\ref{eq:photon_number}) shows that with the given parameters the average photon number per pulse has decreased to 1.12 as early as the third roundtrip. 
Considering that the pulses emitted by the laser exhibit a Poissonian photon number distribution, this number corresponds to a probability of 34.5 $\%$ that multiple photons arrive at the detector simultaneously. 
In the fifth roundtrip, this probability has decreased to 3.9$\%$.\\
As the damage threshold of the APDs is specified with $10^5$ photons per second, there is still space for increasing the input power in order to improve the SNR for higher step numbers. 
In our measurements it is possible to observe step numbers up to 21, while still maintaining a safety margin to the damage threshold. 
In this case, however, considerable multi-photons contributions in early step numbers required additional measurements with lower initial power levels to obtain reliable results for these steps.

\subsection{Reducing losses}
\label{sec:ReducingLosses}
In the previous subsection, we have described how we can balance the multi-photon contributions in the first steps and the SNR in the later steps by adjusting the input power.
Now the focus is on loss minimization during the round trips. Figure \ref{fig:roundtrip} shows the attainable step number depending on the loss factor for three different dynamic ranges of the detectors. We have some freedom in tuning the dynamic range of the detectors by using different initial power levels for high and low step numbers. However, we see that the dominant factor determining the step number is the loss factor. Thus, pushing the setup to a higher number of measurable steps requiring minimising the losses.


For our setup with a freespace-to-fibre based design, the fibre-to-fibre coupling efficiency is the most crucial parameter in the alignment and makes up the biggest share of the overall losses. 

In the alignment process we minimize losses in the first place by adjusting for the best possible mode overlap. 
However, two fibre outcouplings are linked to two incouplings, which results in four different coupling combinations.
As these coupling paths are of different lengths and pass different optical components, it is not realistic to achieve completely equal and at the same time minimal losses for all four paths. 

\begin{figure}[t]
	\centering 
		\includegraphics[width=\columnwidth]{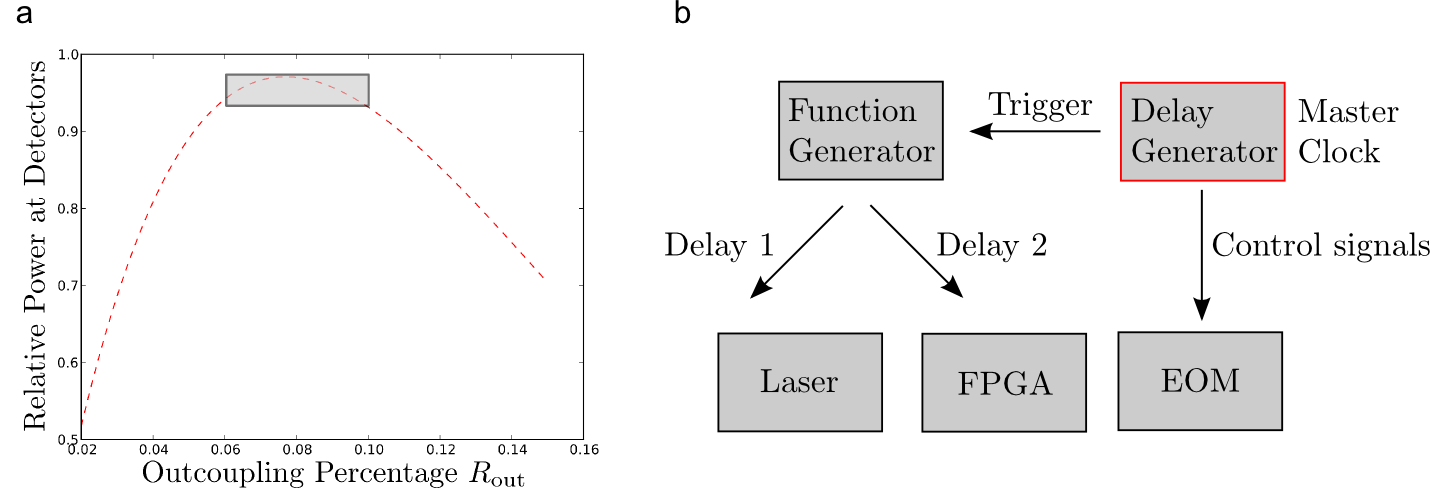}
	\caption{Left figure: The relative power at the detectors depending on the outcoupling percentage $R_\mathrm{out}$ for a step number of 13, the optimal values for reflectivities are marked by the grey box (between 0.06 and 0.10); Right figure: Schematic of the components significant for the timings in the setup and the signals sent between them; FPGA: field-programmable gate array}
\label{fig:Out_Time}
\end{figure}

Another significant loss origin is the partially reflecting beam sampler that couples out a certain proportion of the light in each roundtrip for the read-out of the walker's time evolution. 
The predicament concerning the power at the detectors is that a higher power directed to them and a good SNR will cause at the same time higher losses in each roundtrip.
The optimal outcoupling ratio $R_\mathrm{out}$ for which the power at the detectors is maximal for a certain step number is obtained by finding the maximum of \ref{eq:photon_number} with respect to $R_\mathrm{out}$ and yields $R_\mathrm{out} = \frac{1}{n +1}$.
Figure \ref{fig:Out_Time}a shows the relative photon numbers at the detectors for outcoupling percentages $R_\mathrm{out}$ of 0.02 to 0.15, a roundtrip number $n$ of 13 and values of 0.5 for $L_{\mathrm{RT}}$ and of 0.97 for $L_{\mathrm{BS}}$, while normalizing $P_\mathrm{L}$ to one. 
We see that it is desirable to keep $R_{\mathrm{out}}$ between 0.06 and 0.10 (considering what is technically feasible) in order to get a power level at the detectors that is close to the maximum (see grey shaded range in figure \ref{fig:Out_Time}a).
As we use an outcoupler with a reflectivity of roughly 7$\%$, we are in a region close to the maximum.
\\
The light is coupled into the setup by a partially reflecting incoupler, also introducing losses in each roundtrip when passing through the incoupler. 
As the incoupler affects the initial power linearly and the roundtrip losses exponentially, an ideal incoupler would reflect into the setup just enough intensity for alignment and is optimised for maximal transmission. 
Also considering feasibility, we use an AR-coated $SiO_2$-plate. 
It exhibits an incoupling efficiency of $0.2\%$ and introduces only 3$\%$ loss in a roundtrip.

Summarizing, the adjustment of the given parameters, the initial power and the fibre-to-fibre couplings, as well as the design of optimal in- and outcoupling beam samplers allow for a good loss control in the setup and for an increase of the number of observable steps.

\subsection{Synchronisation}

As our quantum walk relies on the accurate addressing and read-out of the time-multiplexed positions, it requires a proper synchronisation between the laser pulses, the dynamically switching EOM and the detection windows of the field-programmable gate array (FPGA) acting as a time-to-digital converter.
The synchronisation scheme is illustrated in Figure \ref{fig:Out_Time}b: A delay generator acts as the master clock, on the one hand sending a trigger signal with a given frequency to a function generator. 
The function generator has two output channels, one connected to the laser and the other to a field-programmable gate array (FPGA) and allows for adjusting an individual delay for each of them. 
On the other hand, the delay generator delivers the signals that actually control to the application of the high voltage to the EOM's crystal.
\\
This setup allows us to adjust the delay between the laser and the switching pattern applied to the EOM as well as the delay between the laser and the FPGA, making it possible to deliberately switch the desired pulses.

\section{Results} \label{sec:results}

The capability of implementing dynamical coin operations is an essential ingredient for full dynamical control, making the optical feedback loop a highly versatile testbed.
With this ability we can realise dynamically changeable graph structures as reported in Ref.~\cite{elster_quantum_2015}, to which we add here the following three new applications.

First, we show the implementation of the specific example of a quantum walk on a finite graph, which has so far only been implemented in static waveguide arrays with a fixed number of positions for continuous-time quantum walks \cite{meinecke_coherent_2013}.

The preparation of different input states significantly enhances the possibilities in studying the interaction of the internal degree of freedom and the output intensity distribution.
In addition, quantum search algorithms typically start with non-localised initial states for optimal search results \cite{shenvi_quantum_2003, ambainis_coins_2005}. 
Thus, we present, secondly, a scheme for the in-situ preparation of non-localised input states.

Last, we show two experimental realisations of state transfer schemes \cite{zhan_perfect_2014} making use of the dynamical coin operation.

\subsection{Quantum Walk on a Finite Graph} 
\label{subsec:finite}
The experimental implementation of precisely timed switching operations enables us to modify the graph structure on which the quantum walk takes place. 
In the following, we present two examples, where we limit the spread of the walker's wave function to a preset number of sites (see Figure \ref{fig:Kaskade_Timings}), while even more complex structures are possible as well as recently demonstrated in \cite{elster_quantum_2015}.

For an illustrative presentation of the measured results, we will employ two-dimensional chessboard diagrams with the vertical axis corresponding to the step and the horizontal axis to the position number. 
The intensities assigned to the individual positions are colour-coded and traced out over polarization. 

The error bars for all the numerical data presented in this and the following subsections in bar chart representations of single steps are obtained by conducting a Monte Carlo scan which accounts for possible variations of the coupling efficiencies by 1.5$\%$, of the transmission coefficient of the EOM by 2$\%$ and of the coin angle by 0.1$^{\circ}$ as described in detail in \cite{elster_quantum_2015}. 
These errors are much greater than the statistical errors of the measured data and are therefore considered the relevant errors.

The distance measure (Manhattan distance) used in this article is determined by summing up the absolute values of the differences between the experimental and numerical values of the normalized occupation probabilities $P_x$ for all positions $x$ and both polarizations (H and V). 
Its possible values range from 0 to 1.
\begin{eqnarray}
d = \frac{1}{2} \left( \sum_x | P_{H,x}^{(\mathrm{num})} - P_{H,x}^{(\mathrm{exp})}| 
 + \sum_x | P_{V,x}^{(\mathrm{num})} - P_{V,x}^{(\mathrm{exp})}| \right)
\label{eq:distance}
\end{eqnarray}
When presenting the data for several steps we calculated the distance for each step and present the average number for all steps under consideration.
 
\begin{figure}[t]
	\centering 
		\includegraphics[width=\columnwidth]{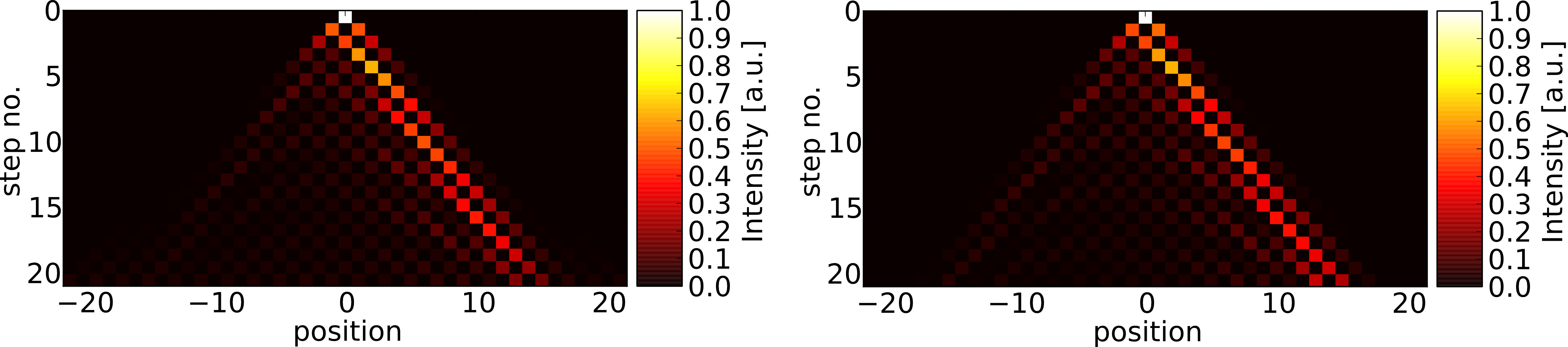}
	\caption{Chessboard representation of the experimental (left) and the numerical data (right) of an unrestricted walk for a horizontally polarized input state; the mean distance between the experimental and numerical data is 0.068 per step (considering steps 1 - 20)}
\label{fig:inf_exp_theo}
\end{figure}

Figure \ref{fig:inf_exp_theo} shows the experimental (left) as well as the corresponding numerical data (right) for an unrestricted walk in which $\hat{C}_\mathrm{QWP}$ can be best reproduced by a QWP set to 36$^\circ$.
We observe that the walker's wave function spreads over a pyramidal pattern of positions with a significant bias towards positive positions. 
This bias is observed when --depending on the coin angle-- launching an input state in horizontal polarisation into the setup, while a vertical input state corresponds to a bias towards negative positions. 
This phenomenon is caused by the phases of $\hat{C}_\mathrm{QWP}$ (see eq. (\ref{eq:C_EOM_T})), which introduces different phases for $| H \rangle$ and $| V \rangle$ (see also \cite{kempe_quantum_2003}).

\begin{figure}[t]
	\centering 
		\includegraphics[width=\columnwidth]{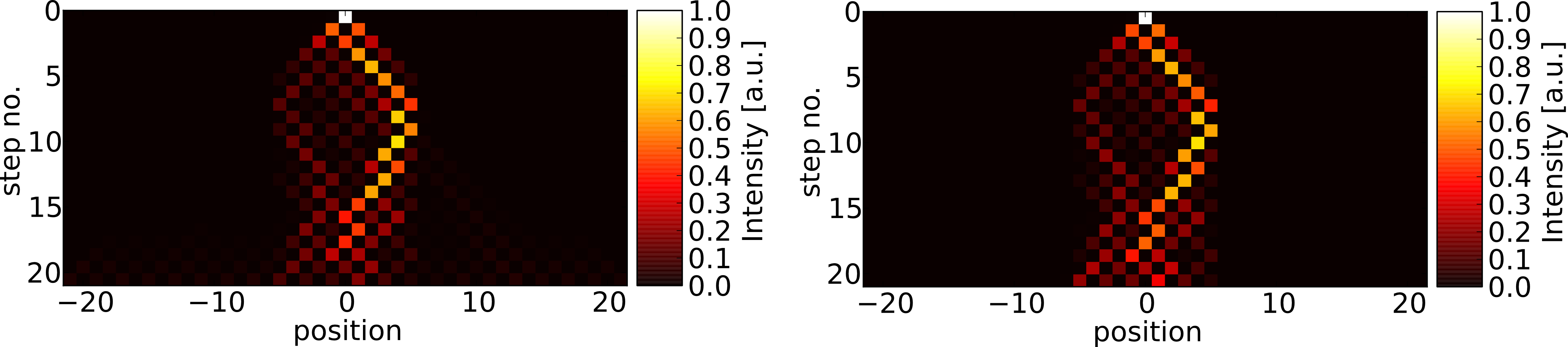}
	\caption{Chessboard representation of the experimental data (left) and the numerical data (right) of a walk on a graph with a maximum of 6 simultaneously occupied positions for a horizontally polarized input state; the mean distance between the experimental and numerical data is 0.086 per step (considering steps 1 - 20)}
\label{fig:5site_exp_theo}
\end{figure}

As pointed out in subsection \ref{subsec:dyna_coin}, we are able to restrict the spread of the walker's wave function by turning the polarisation at the outer positions by 90$^\circ$. 
On the other positions, the coin operation is conducted by a static QWP (compare Figure \ref{fig:Bloch_too_fast}). 
For the first example given here, we show the evolution on a graph with a maximum of 6 simultaneously occupied positions, which is realised by reflections at $x = \pm 5$ (see Figure \ref{fig:5site_exp_theo}).
We observe a good overlap between the experimental and the numerical data with a mean distance of 0.086 per step (taking steps 1 - 20 into account), which confirms that the EOM's operator matrices were worked out correctly (see eq. \ref{eq:C_EOM_U}).

\begin{figure}[t]
	\centering 
		\includegraphics[width=\columnwidth]{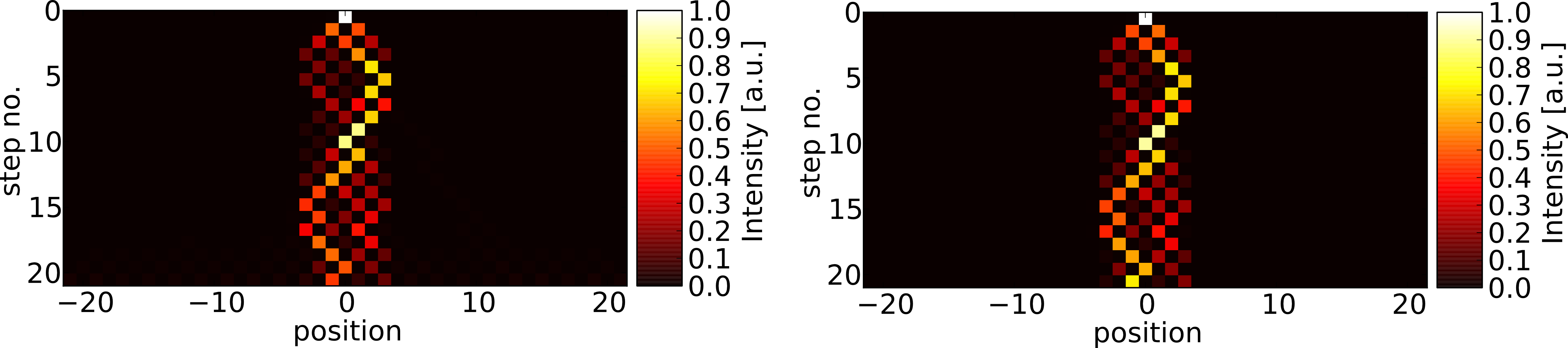}
	\caption{Chessboard representation of the experimental data (left) and the numerical data (right) of a walk on a graph with a maximum of 4 simultaneously occupied positions for a horizontally polarised input state; the mean distance between the experimental and numerical data is 0.052 per step (considering steps 1 - 20)}
\label{fig:3site_exp_theo}
\end{figure}

As the graph is only determined by the switching patterns of the EOM, we can vary the graph size just by reprogramming of the EOM.
Figure \ref{fig:3site_exp_theo} shows a chessboard diagram for an evolution of the walker that is now restricted to a maximum of 4 simultaneously occupied sites (i.e. reflections at $x = \pm 3$).

\begin{figure}[t]
	\centering 
		\includegraphics[width=8cm]{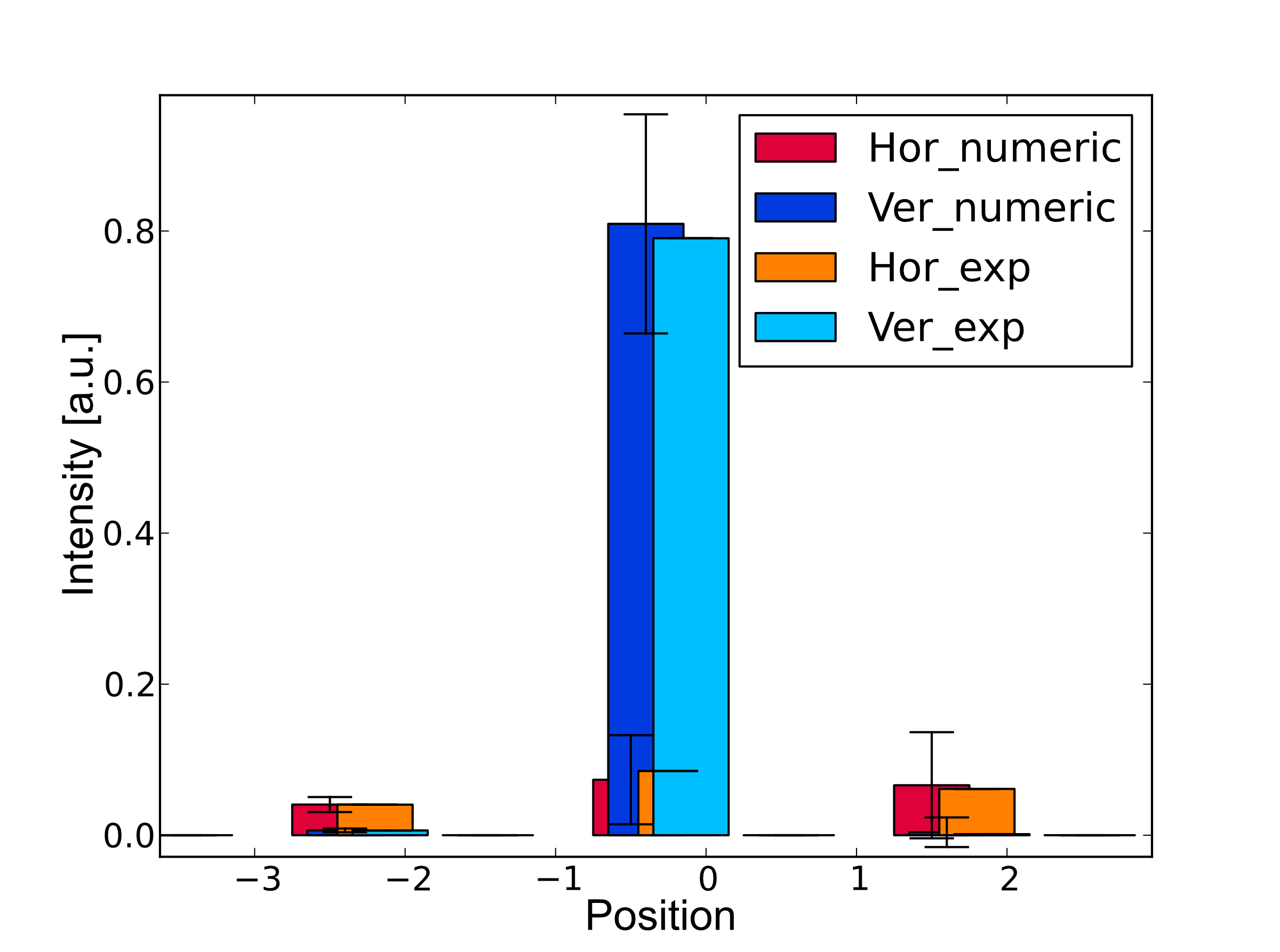}
	\caption{Bar chart representation of the 10th step, the probabilities for the position -2 and 2 are due to experimental imperfections accounted for in the numerics; the data is obtained for a walk on a graph with a maximum of 4 simultaneously occupied positions; the error bars for the numerical data are obtained by Monte Carlo scans, see text for details}
\label{fig:3site_step10}
\end{figure}

Here, we observe an interesting feature of the walk on a finite graph: The walker, initially spread out, is not only reflected at the boundaries, but even exhibits revival (within experimental errors) at the initial position 0 in step 10 (see Figure \ref{fig:3site_step10}). This behavior is predicted by theory for a coherent evolution and shows that the switching operations preserve the coherence of the walker.

\begin{figure}[t]
	\centering 
		\includegraphics[width=\columnwidth]{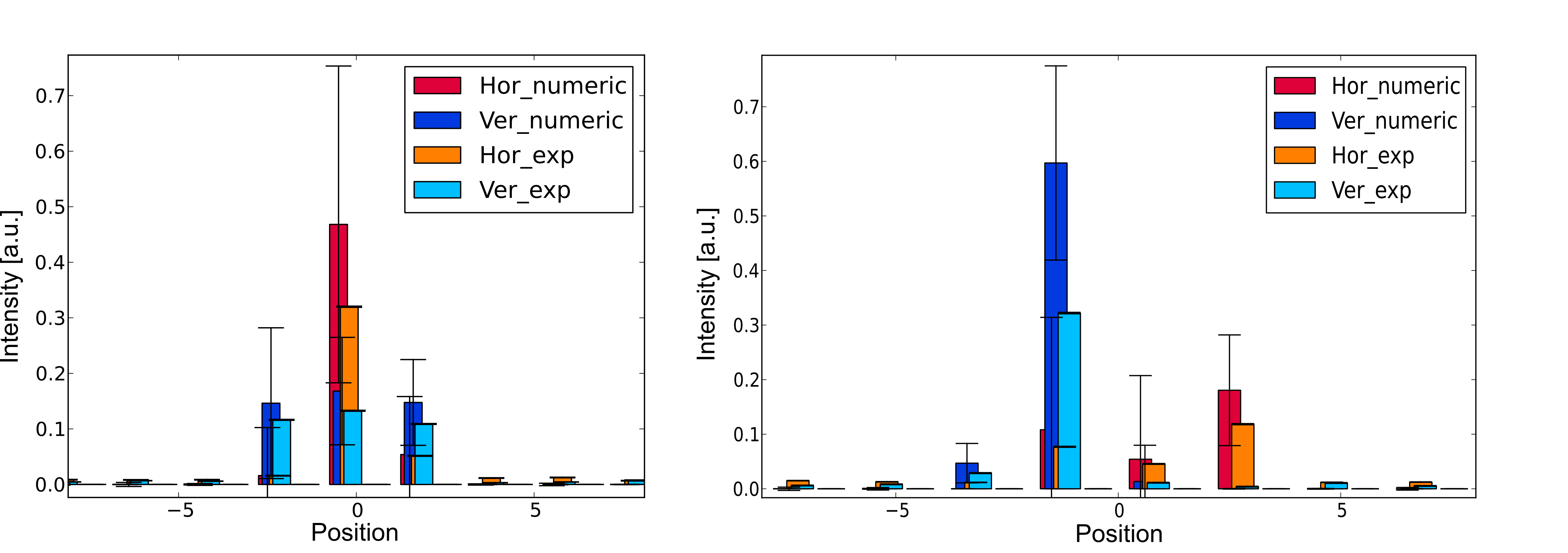}
	\caption{Probability distribution for the occupation of the walker's positions in step 20 (left plot) and 21 (right plot) for experimental and numerical data and both polarizations; the error bars for the numerical data are obtained by Monte Carlo scans, see text for details}
\label{fig:3site_20_21}
\end{figure}

The evolution on the finite graph is reliably recorded for a quantum walk up to step 21. 
In general, losses scale exponentially with the step number, rendering the implementation of quantum walks with high step numbers challenging. 
In order to achieve a good signal-to-noise ratio for the step numbers presented here, we take measurements with two different initial power levels: For the first measurement, we use a neutral-density-filter (ND-filter) with an optical density of 4 which allows for reliable detection in the first steps, however, we do not observe a good SNR for step numbers greater than 13. 
Thus, we perform a second measurement employing a ND-filter with an optical density of 1, which results in an initial photon number that is by a factor of 1000 higher. 
In this case, the average photon number sent to the detectors has decreased to 0.03 in step 8 (the first step to be analysed for the higher power level), while we still obtain reliable results up to step 20 (see Figure \ref{fig:3site_20_21}). 
Step 21 is the first step in which the deviation between the experimental and the numeric data is for the first time greater than the error bars and thus determines the limit in reliably observable step numbers. 
The data shown in Figures \ref{fig:inf_exp_theo}, \ref{fig:5site_exp_theo} and \ref{fig:3site_exp_theo} is thus concatenated from ND4-measurements for step 1 to 8 or 9 and ND1-measurements for the subsequent steps. 
These experiments are conducted with a repetition rate of 50 kHz.

\subsection{State Preparation}	
\label{subsec:state_prep}
\begin{figure}[t]
	\centering 
		\includegraphics[width=\columnwidth]{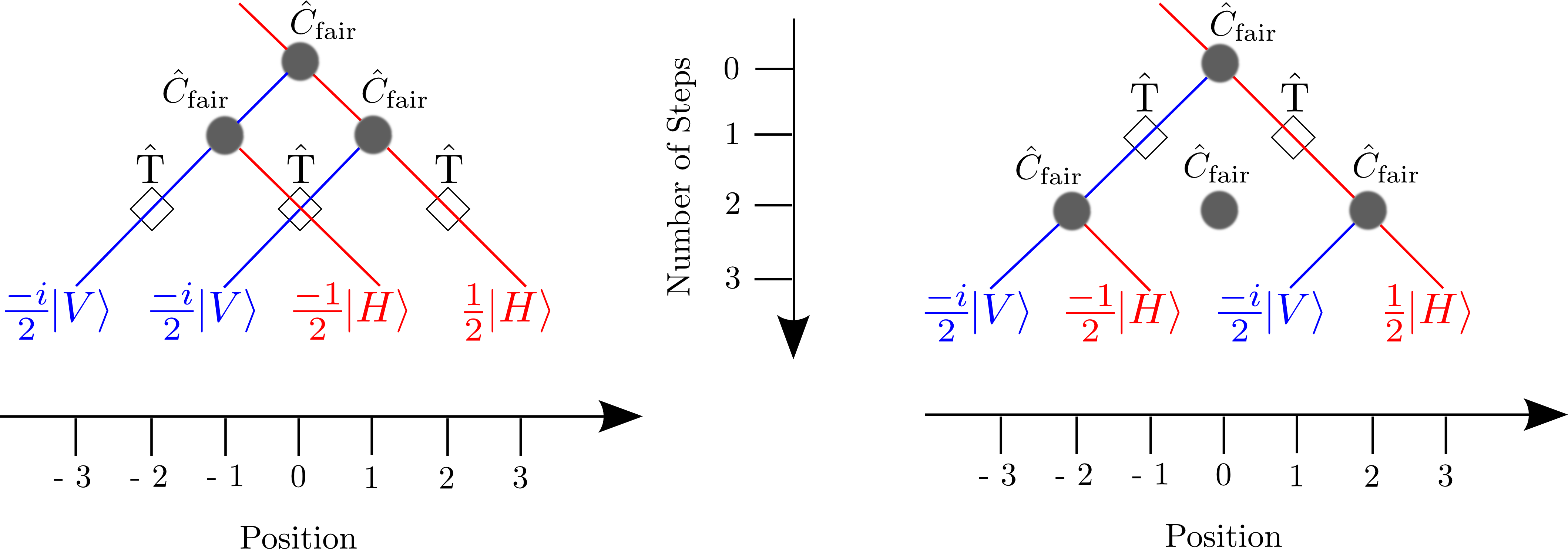}
	\caption{Illustration of the implementation of the states $| \Psi_\mathrm{VVHH} \rangle$ (left figure) and $| \Psi_\mathrm{VHVH} \rangle$ (right figure) with the three-state-switching scheme for an horizontally polarised input state}
\label{fig:State_Prep}
\end{figure}

Most quantum walk experiments including those presented in subsection \ref{subsec:finite} are performed with localised input states. 
However, boson sampling experiments in a time-multiplexing setup \cite{motes_scalable_2014, he_scalable_2016} similar to the presented system rely on initial states occupying several positions.
Quantum search algorithms \cite{shenvi_quantum_2003, ambainis_coins_2005} require in addition a well-defined phase relation of the different positions.
The ''in-situ'' state preparation, i.e. the preparation of different initial states within the quantum walk setup, guarantees phase-stable pulse trains with controlled polarization inherently exhibiting a coherent evolution.
Full dynamical coin control yields the possibility to realise arbitrary passive optical transformations of a fixed state \cite{reck_experimental_1994}, in turn allowing the preparation of arbitrary states.
Two examples of transformations accessible by the EOM used in the present setup are illustrated in Figure \ref{fig:State_Prep}, we use these schemes to prepare two states with four positions, but different polarisation with the three-state-switching scheme. Applying the coin operation $\hat{C}_\mathrm{QWP}$ in the first step and the transmission operation $\hat{T}$ in the second step (case A) results in the walker's two left positions being vertically polarized and the two right positions being horizontally polarized.
The full input state including all phases is then given by $| \Psi_\mathrm{VVHH} \rangle \equiv 1 /2 ~( - i | - 3, V \rangle  - i | - 1,V \rangle -| 1,H \rangle +| 3,H \rangle$). 
When now switching the transmission operation $\hat{T}$ in the first step and $\hat{C}_\mathrm{QWP}$ in the second (case B), the positions -3 and 1 are vertically polarized and the positions -1 and 3 horizontally polarized (input state: $| \Psi_\mathrm{VHVH} \rangle \equiv 1 /2 ~( - i | - 3,V \rangle  - | - 1, H \rangle - i | 1,V \rangle +| 3,H \rangle$).

\begin{figure}[t]
	\centering 
		\includegraphics[width=\columnwidth]{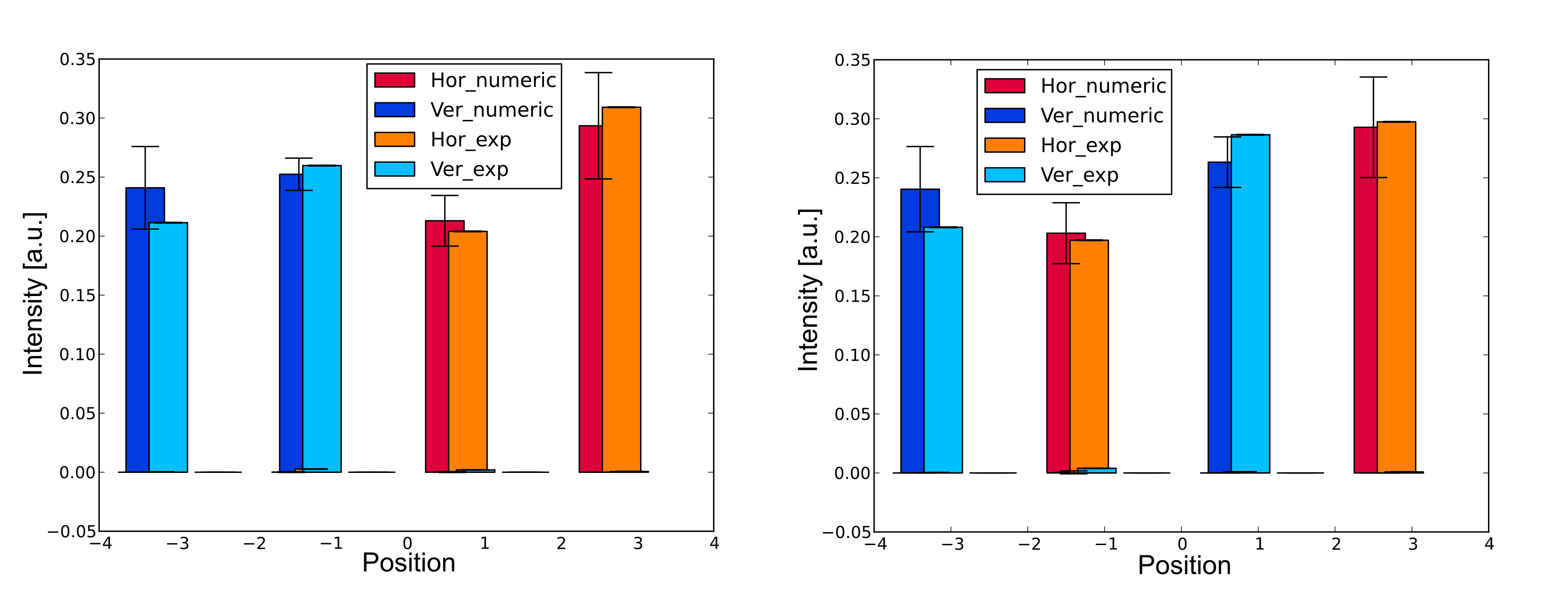}
	\caption{Bar chart representation of the experimental and numerical probabilities for the input states $| \Psi_\mathrm{VVHH} \rangle$ (left) and $| \Psi_\mathrm{VHVH} \rangle$ (right) after their preparation (step 3)}
\label{fig:VVHH_VHVH_bar_chart}
\end{figure}

Figure \ref{fig:VVHH_VHVH_bar_chart} shows the initial probability distribution regarding the positions and polarizations of the these two states in a bar chart representation.
The slightly unequal heights of the four peaks result from the different coupling losses in the setup (different polarisations take different paths) and are accounted for in the numerical simulation.

\begin{figure}[t]
	\centering 
		\includegraphics[width=\columnwidth]{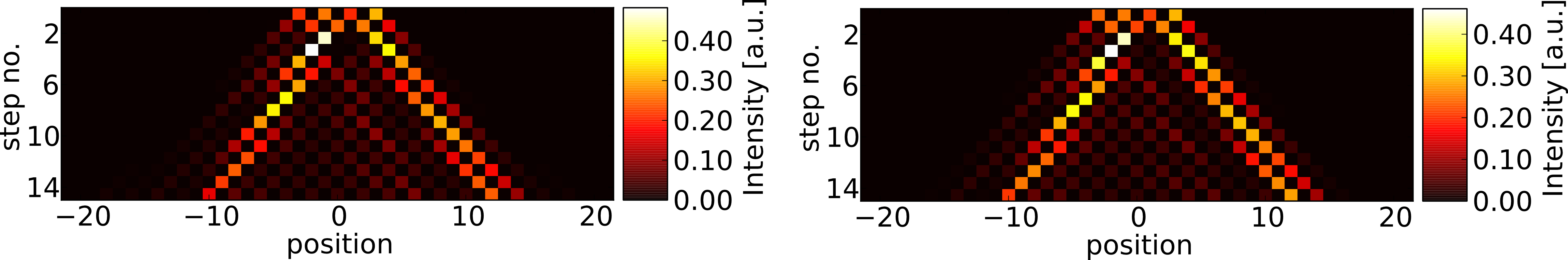}
	\caption{Chessboard representation of the experimental data (left) and the numerical data (right) of an unrestricted walk with the input state $| \Psi_\mathrm{VVHH} \rangle$. Note that the step counting now starts after the 3 steps needed for the state preparation. The observed distance is 0.063, averaged over steps 1 - 14.}
\label{fig:VVHH}
\end{figure}

\begin{figure}[t]
	\centering 
		\includegraphics[width=\columnwidth]{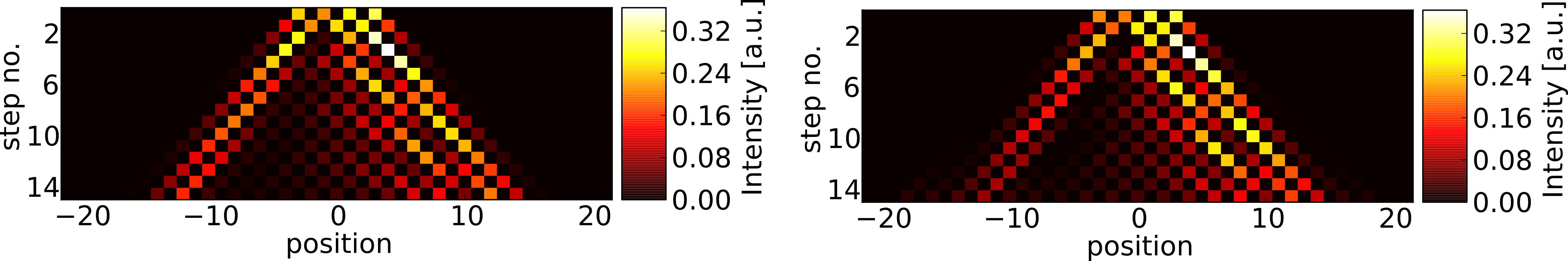}
	\caption{Chessboard representation of the experimental data (left) and the numerical data (right) of an unrestricted walk with the input state $| \Psi_\mathrm{VHVH} \rangle$. The observed distance is 0.091 averaged over steps 1 - 14.}
\label{fig:VHVH}
\end{figure}

For these given input states, we observe the evolution in a balanced walk on an infinite graph (see Fig. \ref{fig:VVHH} for $| \Psi_\mathrm{VVHH} \rangle$ and Fig. \ref{fig:VHVH} for $| \Psi_\mathrm{VHVH} \rangle$).
Here, the experiments are performed with a repetition rate of 110 kHz.

The agreement of the numerical and experimental data proves the lasting coherence of the walker's evolution.

The comparison of the two cases reveals the relevance of the input polarisation: For the $| \Psi_\mathrm{VVHH} \rangle$-state, the walker evolves along two branches which are more distinctively defined than for the $| \Psi_\mathrm{VHVH} \rangle$-state. 
Here in comparison, especially the right branch is much more smeared out leading to a very different intensity distribution in position space.

By reprogramming the EOM switching patterns, the preparation of well-defined input states distributed over a higher number of positions is possible as well.
However, an increasing number of populated initial positions requires an increasing number of roundtrips just for the input state preparation and limits the length of the observable evolution of these states (see role of losses discussed in subsection \ref{subsec:numbers}).

\clearpage

\subsection{State Transfer}	\label{subsec:state_transfer}
\begin{figure}[tb]
	\centering 
		\includegraphics[width=\textwidth]{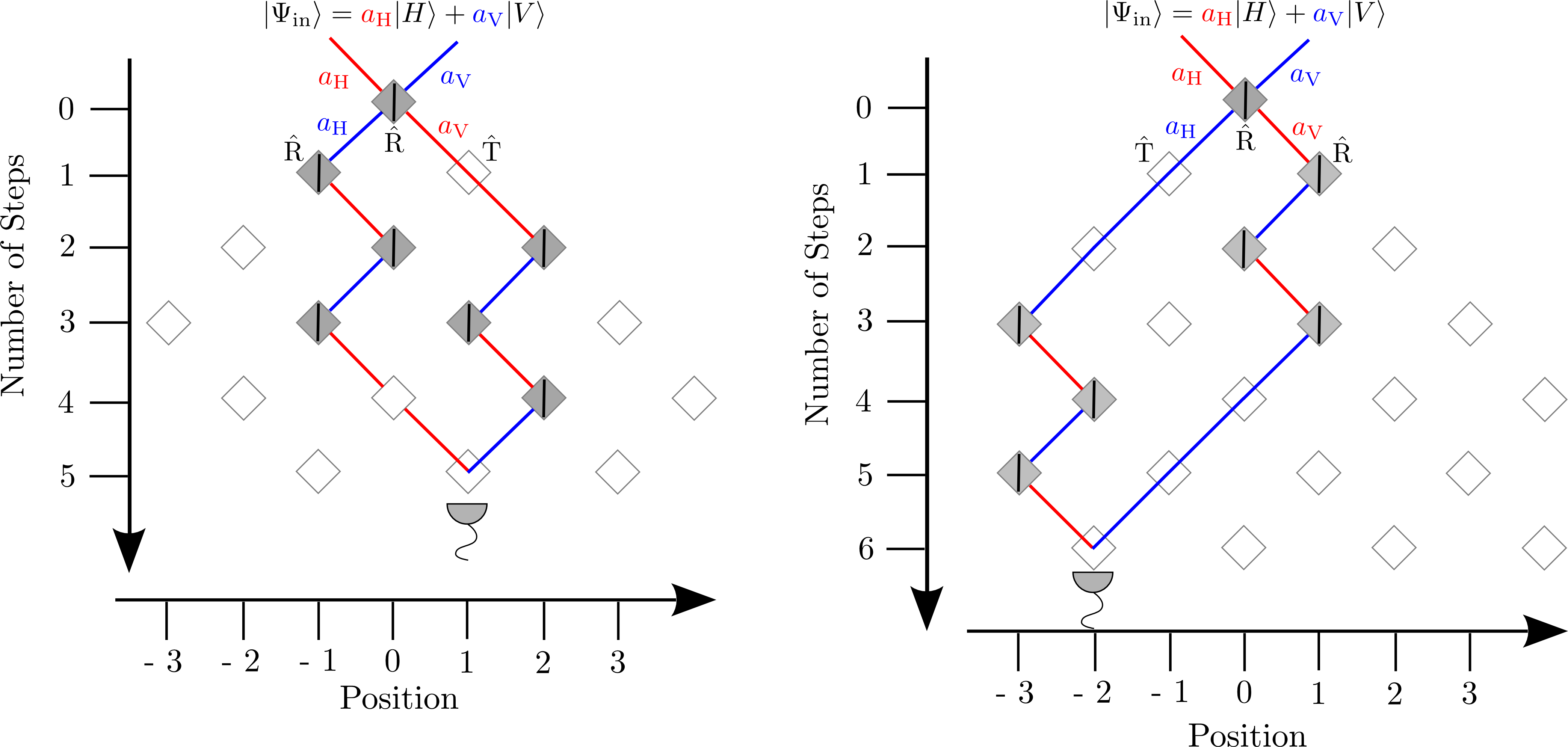}
	\caption{Illustration of the implementation of the 5 step state-transfer scheme (left) and the 6 step state-transfer scheme (right). For both schemes, the phase of i introduced by the reflection operation is compensated for by applying it four times.}
	\label{fig:illu_transfer}
\end{figure}

Zhan \etal in \cite{zhan_perfect_2014} propose a state transfer scheme via discrete time quantum walks which has so far not been realised.
The possibility of dynamical coin control allows for the easy implementation of this protocol in our setup. 
We observe a revival of the walker's initial polarisation state at a chosen position, although it is beforehand spread out in space and undergoes switchings in polarization.
Different protocols transferring the state to different positions can be realised simply by reprogramming the EOM. 
In the following, we show the implementation of 2 exemplary non-trivial schemes. 
The walker's full polarization density matrices are obtained via tomographic measurements \cite{james_measurement_2001}.\\
Figure \ref{fig:illu_transfer}(left) illustrates the switching scheme implemented to transfer a state from position 0 in the 0-th step to position 1 in the 5-th step: An input state of arbitrary polarisation is started at position 0 and reflected and transmitted according to the presented scheme. The phase of i  introduced by the reflection operation is compensated for by applying it four times.

The state transfer continues with a periodicity of 5 steps, while the position of the transferred state oscillates between 0 and 1.

\begin{figure}[t]
	\centering 
		\includegraphics[width=\columnwidth]{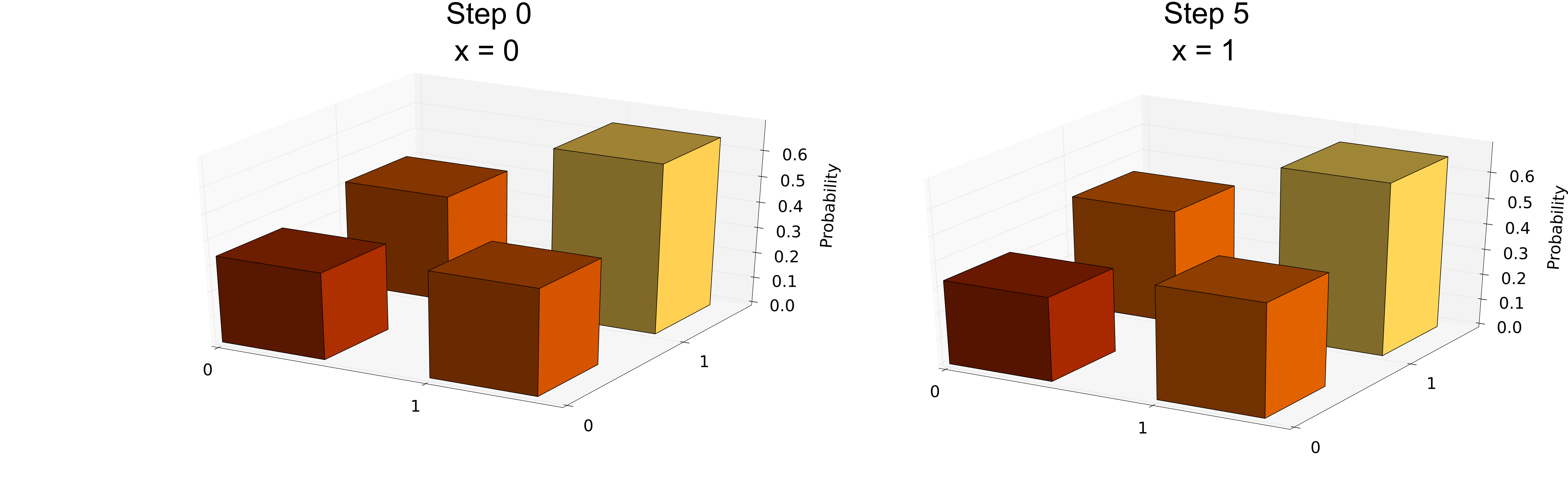}
	\caption{The absolute values of the density matrix of the transferred state in the 0th step (left plot) at the position $x= 0$ and 5th step at the position $x= 1$ (right plot) of a state transfer with a periodicity of 5 steps (scheme illustrated in Figure \ref{fig:illu_transfer}); the observed fidelity $F(\rho_0, \rho_5)$ is 99.4$\%$}
	\label{fig:transfer_matrices_5step}
\end{figure}

Figure \ref{fig:transfer_matrices_5step} shows the evolution of the absolute values of the walker's density matrix from step 0 (position: $x = 0$) to step 5 (position: $x = 1$). The initial density matrix $\rho_0$ is reproduced with a fidelity $F(\rho_0, \rho_5)$ of 99.4$\%$ in step 5 with the density matrices:

\begin{center}
\begin{eqnarray}
\rho_0
= 
\left(
\begin{array}{cc}
0.34 & -0.4 + 0.11 \cdot i \\
-0.4 - 0.11 \cdot i & 0.67
\end{array}
\right)\\
\rho_5
= 
\left(
\begin{array}{cc}
0.33 & -0.42 + 0.14 \cdot i \\
-0.42 - 0.14 \cdot i & 0.67
\end{array}
\right)
\label{eq:transfer_matrices}
\end{eqnarray}
\end{center}

The fidelity $F(\rho_0, \rho_\mathrm{x})$ of a transferred state $\rho_\mathrm{x}$ in step $x$ and the initial state $\rho_0$ is defined as (compare \cite{wang_using_2012}):

\begin{equation}
F(\rho_0, \rho_\mathrm{x}) = \left(\textrm{tr}\left[\sqrt{\sqrt{\rho_0}\rho_\mathrm{x}\sqrt{\rho_0}}\right]\right)^2
\label{eq:fidel}
\end{equation}

In the 10th step, we observe a fidelity $F(\rho_0, \rho_{10})$ to the initial state of 98.7$\%$ and in the 15th step $F(\rho_0, \rho_{15})$ is determined to be 98.3$\%$.\\

\begin{figure}[t]
	\centering 
		\includegraphics[width=\columnwidth]{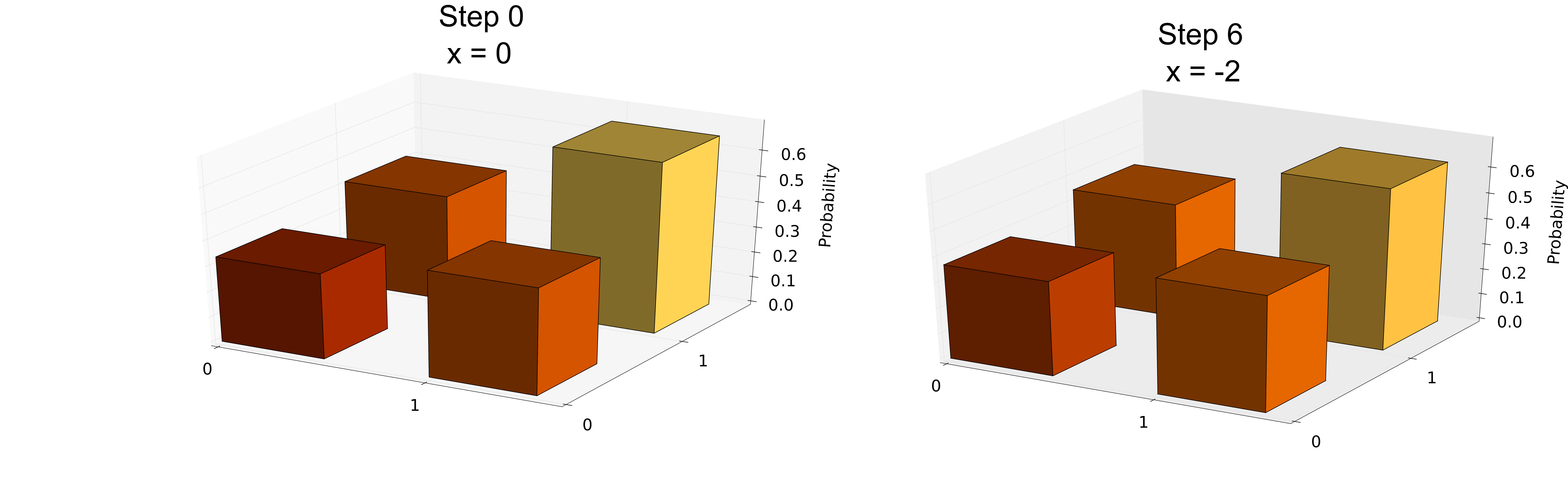}
	\caption{The absolute values of the density matrix of the transferred state in the 0th step (left plot) at the position $x= 0$ and 6th step at the position $x= -2$ (right plot) of a state transfer with a periodicity of 6 steps (scheme illustrated in Figure \ref{fig:illu_transfer}); the observed fidelity $F(\rho_0, \rho_6)$ is 99.5$\%$}
\label{fig:transfer_matrices_6step}
\end{figure}

The 6 step state transfer scheme works analogous to the 5 step scheme, but exhibits a periodicity of 6 steps (see Figure \ref{fig:illu_transfer} (right panel)), with the position of the transferred state oscillating between -2 and 0. We feed in the same input state for the 6 step transfer scheme as for the 5 step transfer scheme.
Figure \ref{fig:transfer_matrices_6step} shows the density matrices $\rho_0$ for step 0 (position: $x = 0$) and $\rho_6$ for step 6 (position: $x = -2$) in this case.

Here, we achieve fidelities of $F(\rho_0, \rho_6) = 99.5\%$ from the initial state to step 6 and of $F(\rho_0, \rho_{12}) = 94.3\%$ from the input state to step 12.
\\
The state transfer scheme is implemented with a repetition rate of 50 kHz.\\
In conclusion, we have shown that state transfer schemes as proposed in \cite{zhan_perfect_2014} can be easily implemented with our setup exhibiting high fidelities. This implementation demonstrates the control guiding of a wave packet across a non-trivial finite topology.

\section{Conclusion} \label{sec:conclusion}

Dynamical coin control in combination with time-multiplexing makes our setup a highly versatile model system for various quantum walk applications: To start with, it allows for implementing a rich variety of graph structures, such as percolating graphs \cite{elster_quantum_2015}. 
In this work, we presented the experimental realisation of two finite graphs restricting the spread of the walker to 4 and 6 simultaneously occupied positions without changing the setup geometry. 
Here, the walker, though spread out over several positions during the walk, exhibits revival at the initial position.\\
We have demonstrated another viable application by the in-situ preparation
of different non-localised input states. 
This possibility allows us to study the influence of the initial polarisation on the evolution of the walker's wave function without losing coherence.
Furthermore, it is even possibile to combine the input state preparation in the first steps with the modification of the underlying graph for the further propagation.
\\
Eventually, we implemented state transfer schemes according to \cite{zhan_perfect_2014} to two different positions with our DTQW-setup. 
We found the full polarization density matrix of the initial state with a high fidelity at a defined receiving positions for a periodicity of either 5 or 6 steps.
\\
Summarizing, we have demonstrated the full dynamic control over the quantum walk evolution on complex graph structures.
As an outlook, the dimensionality of the graph can be extended to a second dimension \cite{schreiber_2d_2012} in order to investigate non-trivial graphs in higher dimensions \cite{kollar_percolation_2014}.
To further increase complexity, a full control over the input states can be established including compatible well-tailored single photon sources of non-classical states \cite{eckstein_highly_2011, harder_optimized_2013} to study the evolution of non-classical properties on a complex dynamic graph. 
An additional direction of enhancement is overcoming losses by amplification \cite{regensburger_photon_2011}.

\section{Acknowledgements} 

J.~N. and I.~J.\ have been partially supported by the Czech Science foundation (GA{\v C}R) project number 16-09824S, A.~G., and I.~J.\ from No.\ GA{\v C}R 13-33906S, and A.~G.\ from T\'AMOP-4.2.4.A/2-11/1-2012-0001 ``National Excellence Program'' of Hungary.\\
T.N., F.E., S.B. and C.S. acknowledge funding by the DFG (Deutsche Forschungsgemeinschaft) via the Gottfried Wilhelm Leibniz-Preis.\\
This work has received funding from the European Union’s Horizon 2020 research and innovation programme under the QUCHIP project GA no. 641039.

\clearpage

\bibliographystyle{unsrt}
\bibliography{NewLib}

\begin{thebibliography}{10}

\bibitem{aharonov_quantum_1993}
Y.~Aharonov, L.~Davidovich, and N.~Zagury.
\newblock Quantum random walks.
\newblock {\em Physical Review A}, 48(2):1687--1690, August 1993.

\bibitem{kempe_quantum_2003}
J~Kempe.
\newblock Quantum random walks: {An} introductory overview.
\newblock {\em Contemporary Physics}, 44(4):307--327, 2003.

\bibitem{plenio_dephasing-assisted_2008}
M.~B. Plenio and S.~F. Huelga.
\newblock Dephasing-assisted transport: quantum networks and biomolecules.
\newblock {\em New Journal of Physics}, 10(11):113019, November 2008.

\bibitem{mohseni_environment-assisted_2008}
Masoud Mohseni, Patrick Rebentrost, Seth Lloyd, and Alán Aspuru-Guzik.
\newblock Environment-assisted quantum walks in photosynthetic energy transfer.
\newblock {\em The Journal of Chemical Physics}, 129(17):174106, November 2008.

\bibitem{ahlbrecht_molecular_2012}
Andre Ahlbrecht, Andrea Alberti, Dieter Meschede, Volkher~B. Scholz, Albert~H.
  Werner, and Reinhard~F. Werner.
\newblock Molecular binding in interacting quantum walks.
\newblock {\em New Journal of Physics}, 14(7):073050, July 2012.

\bibitem{shikano_discrete-time_2014}
Yutaka Shikano, Tatsuaki Wada, and Junsei Horikawa.
\newblock Discrete-time quantum walk with feed-forward quantum coin.
\newblock {\em Scientific Reports}, 4, March 2014.

\bibitem{kitagawa_exploring_2010}
Takuya Kitagawa, Mark Rudner, Erez Berg, and Eugene Demler.
\newblock Exploring topological phases with quantum walks.
\newblock {\em Physical Review A}, 82(3), 2010.

\bibitem{kitagawa_topological_2012}
Takuya Kitagawa.
\newblock Topological phenomena in quantum walks: elementary introduction to
  the physics of topological phases.
\newblock {\em Quantum Information Processing}, 11(5):1107--1148, August 2012.

\bibitem{asboth_symmetries_2012}
J.~K. Asbóth.
\newblock Symmetries, topological phases, and bound states in the
  one-dimensional quantum walk.
\newblock {\em Physical Review B}, 86(19):195414, November 2012.

\bibitem{lovett_universal_2010}
Neil~B. Lovett, Sally Cooper, Matthew Everitt, Matthew Trevers, and Viv Kendon.
\newblock Universal quantum computation using the discrete-time quantum walk.
\newblock {\em Physical Review A}, 81(4):042330, April 2010.

\bibitem{childs_universal_2009}
Andrew~M. Childs.
\newblock Universal {Computation} by {Quantum} {Walk}.
\newblock {\em Physical Review Letters}, 102(18):180501, 2009.

\bibitem{childs_universal_2013}
Andrew~M. Childs, David Gosset, and Zak Webb.
\newblock Universal {Computation} by {Multiparticle} {Quantum} {Walk}.
\newblock {\em Science}, 339(6121):791--794, February 2013.

\bibitem{du_experimental_2003}
Jiangfeng Du, Hui Li, Xiaodong Xu, Mingjun Shi, Jihui Wu, Xianyi Zhou, and
  Rongdian Han.
\newblock Experimental implementation of the quantum random-walk algorithm.
\newblock {\em Physical Review A}, 67(4):042316, April 2003.

\bibitem{ryan_experimental_2005}
C.~A. Ryan, M.~Laforest, J.~C. Boileau, and R.~Laflamme.
\newblock Experimental implementation of a discrete-time quantum random walk on
  an {NMR} quantum-information processor.
\newblock {\em Physical Review A}, 72(6):062317, 2005.

\bibitem{schmitz_quantum_2009}
H.~Schmitz, R.~Matjeschk, Ch. Schneider, J.~Glueckert, M.~Enderlein, T.~Huber,
  and T.~Schaetz.
\newblock Quantum {Walk} of a {Trapped} {Ion} in {Phase} {Space}.
\newblock {\em Physical Review Letters}, 103(9):090504, August 2009.

\bibitem{zahringer_realization_2010}
F.~Zähringer, G.~Kirchmair, R.~Gerritsma, E.~Solano, R.~Blatt, and C.~F. Roos.
\newblock Realization of a {Quantum} {Walk} with {One} and {Two} {Trapped}
  {Ions}.
\newblock {\em Physical Review Letters}, 104(10):100503, 2010.

\bibitem{karski_quantum_2009}
Michal Karski, Leonid Förster, Jai-Min Choi, Andreas Steffen, Wolfgang Alt,
  Dieter Meschede, and Artur Widera.
\newblock Quantum {Walk} in {Position} {Space} with {Single} {Optically}
  {Trapped} {Atoms}.
\newblock {\em Science}, 325(5937):174--177, July 2009.

\bibitem{genske_electric_2013}
Maximilian Genske, Wolfgang Alt, Andreas Steffen, Albert~H. Werner, Reinhard~F.
  Werner, Dieter Meschede, and Andrea Alberti.
\newblock Electric {Quantum} {Walks} with {Individual} {Atoms}.
\newblock {\em Physical Review Letters}, 110(19):190601, 2013.

\bibitem{bouwmeester_optical_1999}
D.~Bouwmeester, I.~Marzoli, G.~Karman, W.~Schleich, and J.~Woerdman.
\newblock Optical {Galton} board.
\newblock {\em Physical Review A}, 61(1):1--9, December 1999.

\bibitem{do_experimental_2005}
Binh Do, Michael~L. Stohler, Sunder Balasubramanian, Daniel~S. Elliott,
  Christopher Eash, Ephraim Fischbach, Michael~A. Fischbach, Arthur Mills, and
  Benjamin Zwickl.
\newblock Experimental realization of a quantum quincunx by use of linear
  optical elements.
\newblock {\em Journal of the Optical Society of America B}, 22(2):499--504,
  February 2005.

\bibitem{broome_discrete_2010}
M.~A. Broome, A.~Fedrizzi, B.~P. Lanyon, I.~Kassal, A.~Aspuru-Guzik, and A.~G.
  White.
\newblock Discrete {Single}-{Photon} {Quantum} {Walks} with {Tunable}
  {Decoherence}.
\newblock {\em Physical Review Letters}, 104(15):153602, April 2010.

\bibitem{regensburger_photon_2011}
Alois Regensburger, Christoph Bersch, Benjamin Hinrichs, Georgy Onishchukov,
  Andreas Schreiber, Christine Silberhorn, and Ulf Peschel.
\newblock Photon {Propagation} in a {Discrete} {Fiber} {Network}: {An}
  {Interplay} of {Coherence} and {Losses}.
\newblock {\em Physical Review Letters}, 107(23):233902, December 2011.

\bibitem{cardano_quantum_2015}
Filippo Cardano, Francesco Massa, Hammam Qassim, Ebrahim Karimi, Sergei
  Slussarenko, Domenico Paparo, Corrado~de Lisio, Fabio Sciarrino, Enrico
  Santamato, Robert~W. Boyd, and Lorenzo Marrucci.
\newblock Quantum walks and wavepacket dynamics on a lattice with twisted
  photons.
\newblock {\em Science Advances}, 1(2):e1500087, March 2015.

\bibitem{xue_observation_2014}
Peng Xue, Hao Qin, Bao Tang, and Barry~C. Sanders.
\newblock Observation of quasiperiodic dynamics in a one-dimensional quantum
  walk of single photons in space.
\newblock {\em New Journal of Physics}, 16(5):053009, 2014.

\bibitem{perets_realization_2008}
Hagai Perets, Yoav Lahini, Francesca Pozzi, Marc Sorel, Roberto Morandotti, and
  Yaron Silberberg.
\newblock Realization of {Quantum} {Walks} with {Negligible} {Decoherence} in
  {Waveguide} {Lattices}.
\newblock {\em Physical Review Letters}, 100(17):1--4, May 2008.

\bibitem{bromberg_quantum_2009}
Yaron Bromberg, Yoav Lahini, Roberto Morandotti, and Yaron Silberberg.
\newblock Quantum and {Classical} {Correlations} in {Waveguide} {Lattices}.
\newblock {\em Physical Review Letters}, 102(25):253904, June 2009.

\bibitem{peruzzo_quantum_2010}
Alberto Peruzzo, Mirko Lobino, Jonathan C.~F. Matthews, Nobuyuki Matsuda,
  Alberto Politi, Konstantinos Poulios, Xiao-Qi Zhou, Yoav Lahini, Nur Ismail,
  Kerstin Wörhoff, Yaron Bromberg, Yaron Silberberg, Mark~G. Thompson, and
  Jeremy~L. OBrien.
\newblock Quantum {Walks} of {Correlated} {Photons}.
\newblock {\em Science}, 329(5998):1500--1503, September 2010.

\bibitem{owens_two-photon_2011}
J.~O. Owens, M.~A. Broome, D.~N. Biggerstaff, M.~E. Goggin, A.~Fedrizzi,
  T.~Linjordet, M.~Ams, G.~D. Marshall, J.~Twamley, M.~J. Withford, and A.~G.
  White.
\newblock Two-photon quantum walks in an elliptical direct-write waveguide
  array.
\newblock {\em New Journal of Physics}, 13(7):075003, July 2011.

\bibitem{sansoni_two-particle_2012}
Linda Sansoni, Fabio Sciarrino, Giuseppe Vallone, Paolo Mataloni, Andrea
  Crespi, Roberta Ramponi, and Roberto Osellame.
\newblock Two-{Particle} {Bosonic}-{Fermionic} {Quantum} {Walk} via
  {Integrated} {Photonics}.
\newblock {\em Physical Review Letters}, 108(1):010502, January 2012.

\bibitem{di_giuseppe_einstein-podolsky-rosen_2013}
G.~Di~Giuseppe, L.~Martin, A.~Perez-Leija, R.~Keil, F.~Dreisow, S.~Nolte,
  A.~Szameit, A.~F. Abouraddy, D.~N. Christodoulides, and B.~E.~A. Saleh.
\newblock Einstein-{Podolsky}-{Rosen} {Spatial} {Entanglement} in {Ordered} and
  {Anderson} {Photonic} {Lattices}.
\newblock {\em Physical Review Letters}, 110(15):150503, April 2013.

\bibitem{crespi_anderson_2013}
Andrea Crespi, Roberto Osellame, Roberta Ramponi, Vittorio Giovannetti, Rosario
  Fazio, Linda Sansoni, Francesco De~Nicola, Fabio Sciarrino, and Paolo
  Mataloni.
\newblock Anderson localization of entangled photons in an integrated quantum
  walk.
\newblock {\em Nature Photonics}, 7(4):322--328, April 2013.

\bibitem{meinecke_coherent_2013}
J.~D.~A. Meinecke, K.~Poulios, A.~Politi, J.~C.~F. Matthews, A.~Peruzzo,
  N.~Ismail, K.~Wörhoff, J.~L. O’Brien, and M.~G. Thompson.
\newblock Coherent time evolution and boundary conditions of two-photon quantum
  walks in waveguide arrays.
\newblock {\em Physical Review A}, 88(1):012308, July 2013.

\bibitem{poulios_quantum_2014}
Konstantinos Poulios, Robert Keil, Daniel Fry, Jasmin~D.A. Meinecke,
  Jonathan~C.F. Matthews, Alberto Politi, Mirko Lobino, Markus Gräfe, Matthias
  Heinrich, Stefan Nolte, Alexander Szameit, and Jeremy~L. O’Brien.
\newblock Quantum {Walks} of {Correlated} {Photon} {Pairs} in
  {Two}-{Dimensional} {Waveguide} {Arrays}.
\newblock {\em Physical Review Letters}, 112(14):143604, April 2014.

\bibitem{schreiber_photons_2010}
A.~Schreiber, K.~N. Cassemiro, V.~Potoček, A.~Gábris, P.~J. Mosley,
  E.~Andersson, I.~Jex, and Ch. Silberhorn.
\newblock Photons {Walking} the {Line}: {A} {Quantum} {Walk} with {Adjustable}
  {Coin} {Operations}.
\newblock {\em Physical Review Letters}, 104(5):050502, February 2010.

\bibitem{schreiber_decoherence_2011}
A.~Schreiber, K.~N. Cassemiro, V.~Potoček, A.~Gábris, I.~Jex, and Ch.
  Silberhorn.
\newblock Decoherence and {Disorder} in {Quantum} {Walks}: {From} {Ballistic}
  {Spread} to {Localization}.
\newblock {\em Physical Review Letters}, 106(18):180403, 2011.

\bibitem{schreiber_2d_2012}
Andreas Schreiber, Aurél Gábris, Peter~P. Rohde, Kaisa Laiho, Martin
  Štefaňák, Václav Potoček, Craig Hamilton, Igor Jex, and Christine
  Silberhorn.
\newblock A 2d {Quantum} {Walk} {Simulation} of {Two}-{Particle} {Dynamics}.
\newblock {\em Science}, 336(6077):55--58, April 2012.

\bibitem{elster_quantum_2015}
Fabian Elster, Sonja Barkhofen, Thomas Nitsche, Jaroslav Novotný, Aurél
  Gábris, Igor Jex, and Christine Silberhorn.
\newblock Quantum walk coherences on a dynamical percolation graph.
\newblock {\em Scientific Reports}, 5:13495, August 2015.

\bibitem{zhan_perfect_2014}
Xiang Zhan, Hao Qin, Zhi-hao Bian, Jian Li, and Peng Xue.
\newblock Perfect state transfer and efficient quantum routing: {A}
  discrete-time quantum-walk approach.
\newblock {\em Physical Review A}, 90(1):012331, July 2014.

\bibitem{xue_trapping_2014}
Peng Xue, Hao Qin, and Bao Tang.
\newblock Trapping photons on the line: controllable dynamics of a quantum
  walk.
\newblock {\em Scientific Reports}, 4, April 2014.

\bibitem{shenvi_quantum_2003}
Neil Shenvi, Julia Kempe, and K.~Birgitta Whaley.
\newblock Quantum random-walk search algorithm.
\newblock {\em Physical Review A}, 67(5):052307, May 2003.

\bibitem{ambainis_coins_2005}
Andris Ambainis, Julia Kempe, and Alexander Rivosh.
\newblock Coins {Make} {Quantum} {Walks} {Faster}.
\newblock In {\em Proceedings of the {Sixteenth} {Annual} {ACM}-{SIAM}
  {Symposium} on {Discrete} {Algorithms}}, {SODA} '05, pages 1099--1108,
  Philadelphia, PA, USA, 2005. Society for Industrial and Applied Mathematics.

\bibitem{motes_scalable_2014}
Keith~R. Motes, Alexei Gilchrist, Jonathan~P. Dowling, and Peter~P. Rohde.
\newblock Scalable {Boson} {Sampling} with {Time}-{Bin} {Encoding} {Using} a
  {Loop}-{Based} {Architecture}.
\newblock {\em Physical Review Letters}, 113(12):120501, September 2014.

\bibitem{he_scalable_2016}
Yu~He, Zu-En Su, He-Liang Huang, Xing Ding, Jian Qin, Can Wang, S.~Unsleber,
  Chao Chen, Hui Wang, Yu-Ming He, Xi-Lin Wang, Christian Schneider, Martin
  Kamp, Sven Höfling, Chao-Yang Lu, and Jian-Wei Pan.
\newblock Scalable boson sampling with a single-photon device.
\newblock {\em arXiv:1603.04127 [cond-mat, physics:quant-ph]}, March 2016.
\newblock arXiv: 1603.04127.

\bibitem{reck_experimental_1994}
Michael Reck, Anton Zeilinger, Herbert~J. Bernstein, and Philip Bertani.
\newblock Experimental realization of any discrete unitary operator.
\newblock {\em Physical Review Letters}, 73(1):58--61, July 1994.

\bibitem{james_measurement_2001}
Daniel F.~V. James, Paul~G. Kwiat, William~J. Munro, and Andrew~G. White.
\newblock Measurement of qubits.
\newblock {\em Physical Review A}, 64(5):052312, October 2001.

\bibitem{wang_using_2012}
Ying-Dan Wang and Aashish~A. Clerk.
\newblock Using {Interference} for {High} {Fidelity} {Quantum} {State}
  {Transfer} in {Optomechanics}.
\newblock {\em Physical Review Letters}, 108(15):153603, April 2012.

\bibitem{kollar_percolation_2014}
B.~Kollár, J.~Novotný, T.~Kiss, and I.~Jex.
\newblock Percolation induced effects in two-dimensional coined quantum walks:
  analytic asymptotic solutions.
\newblock {\em New Journal of Physics}, 16(2):023002, February 2014.

\bibitem{eckstein_highly_2011}
Andreas Eckstein, Andreas Christ, Peter~J. Mosley, and Christine Silberhorn.
\newblock Highly {Efficient} {Single}-{Pass} {Source} of {Pulsed}
  {Single}-{Mode} {Twin} {Beams} of {Light}.
\newblock {\em Physical Review Letters}, 106(1):013603, January 2011.

\bibitem{harder_optimized_2013}
Georg Harder, Vahid Ansari, Benjamin Brecht, Thomas Dirmeier, Christoph
  Marquardt, and Christine Silberhorn.
\newblock An optimized photon pair source for quantum circuits.
\newblock {\em Optics Express}, 21(12):13975, June 2013.

\end{thebibliography}

\end{document}